\pdfoutput=1
\DocumentMetadata{} 
\documentclass[sigconf,natbib=true,anonymous=false]{acmart}

\AtBeginDocument{%
  \providecommand\BibTeX{{%
    \normalfont B\kern-0.5em{\scshape i\kern-0.25em b}\kern-0.8em\TeX}}}

\copyrightyear{2026}
\acmYear{2026}
\setcopyright{cc}
\setcctype{by}
\acmConference[CIKM '26]{Proceedings of the 35th ACM International Conference on Information and Knowledge Management}{November 07--11, 2026}{Rome, Italy}
\acmBooktitle{Proceedings of the 35th ACM International Conference on Information and Knowledge Management (CIKM '26), November 07--11, 2026, Rome, Italy}
\acmDOI{10.1145/3799682.3840943}
\acmISBN{979-8-4007-2539-5/2026/11}

\usepackage{placeins} 
\usepackage{tcolorbox}       
\usepackage{listings}        
\usepackage{soul}

\usepackage{balance}
\usepackage{pifont}
\usepackage{amsfonts, amsthm, amsmath}
\usepackage[ruled,vlined,linesnumbered]{algorithm2e}
\usepackage{color, colortbl}
\usepackage{enumitem,multirow,graphicx,subcaption,multicol,lipsum,float,adjustbox,makecell}
\newlength{\textfloatsepsave}  
\usepackage{cleveref}
\usepackage{algorithmicx}
\usepackage[page]{appendix} 
\crefformat{section}{\S#2#1#3}
\crefformat{subsection}{\S#2#1#3}
\crefformat{subsubsection}{\S#2#1#3}
\usepackage[normalem]{ulem}
\useunder{\uline}{\ul}{}

\usepackage{tabularx}
\usepackage{ragged2e}
\newcolumntype{L}{>{\RaggedRight\arraybackslash}X}

\usepackage{threeparttable}

\DeclareMathOperator*{\argmax}{argmax}

\begin{document}


\title{Profiling What Matters: Context-Aware Item Profiles from Large-Scale Metadata for LLM Recommenders}

\begin{abstract}
While Large Language Models (LLMs) have significantly advanced reranking in recommendation, effectively leveraging item-side information remains challenging. 
Real-world items are described by vast, heterogeneous, and unstructured metadata, where decision-relevant signals are often implicit, noisy, or buried in long descriptions. 
Moreover, feature salience is highly context-dependent, varying not only across items but also across users. 
Existing methods often rely on item titles, fixed attributes, or static item summaries, which limit personalized and fine-grained item understanding.
To bridge this gap, we propose \proposed, a user context-aware item profiling framework for LLM-based reranking.
\proposed first structures raw metadata and reviews into objective features and subjective traits, and employs a lightweight profiler to select the most relevant information for each user–item pair with limited serving-time overhead.
The resulting profiles are concise and context-specific, providing relevant item-side evidence for the LLM's ranking decision.
Experiments show that \proposed consistently improves LLM-based reranking, highlighting the importance of item profiling that effectively exploits vast item-side information.
Our implementation is available at: \repolink .



\end{abstract}

\vspace{-6cm}


\vspace{-2cm}
\newcommand{\proposed}{CAIRO\xspace}
\newcommand{\challenge}[1]{\textbf{(C#1)}}
\newcommand{\repolink}{\href{https://github.com/rgbeing/CAIRO}{https://github.com/rgbeing/CAIRO}}

\newcommand{\smallsection}[1]{{\vspace{0.03in} \noindent \bf {#1}}}


\author{Dojun Hwang}
\affiliation{
   \institution{Korea University}
   \city{Seoul}
   \country{Republic of Korea}}
\email{dojun2006@korea.ac.kr}

\author{Seunghan Lee}
\affiliation{
   \institution{Korea University}
   \city{Seoul}
   \country{Republic of Korea}}
\email{seunghanlee@korea.ac.kr}

\author{Cheonyoung Park}
\affiliation{
   \institution{KT Corporation}
   \city{Seoul}
   \country{Republic of Korea}}
\email{park.cheonyoung@kt.com}

\author{Sara Yu}
\affiliation{
   \institution{KT Corporation}
   \city{Seoul}
   \country{Republic of Korea}}
\email{sara.yu@kt.com}

\author{SeongKu Kang}
\affiliation{
    \institution{Korea University}
    \city{Seoul}
    \country{Republic of Korea}
}
\authornote{Corresponding author.}
\email{seongkukang@korea.ac.kr}

\maketitle
\vspace{-0.2cm}
\section{Introduction}

\begin{figure}[t]
  \centering
  \hspace*{-0.2cm}
  \includegraphics[width=1.05\columnwidth]{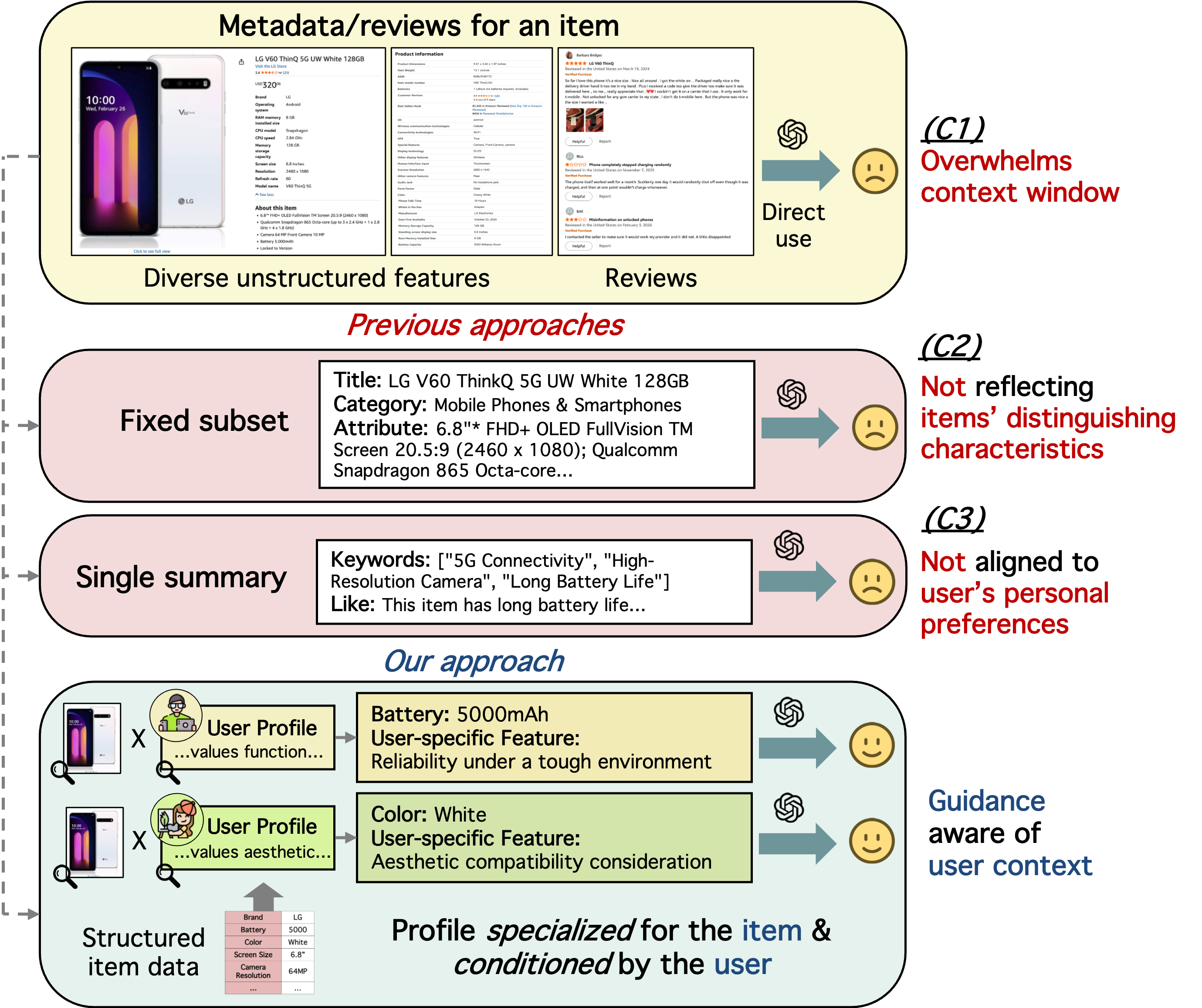}
  \vspace{-3mm}
  \caption{A conceptual comparison of item description exploitation between previous approaches and our approach.}
  \label{fig:concept}
  \vspace{-0.05cm}
\end{figure}

Recently, recommender systems have been embracing a paradigm shift driven by Large Language Models (LLMs).
Leveraging extensive world knowledge and reasoning capabilities, LLMs have successfully addressed the data sparsity issue of traditional ID-based methods by generating rich semantic representations of users and items \cite{recsys-llmera}. 
Within standard multi-stage ranking pipelines \cite{mrs-reranking}, LLMs are widely adopted in the final reranking stage \cite{llamaRec, llm-reranking-1, llm4rerank, corank}.
Provided with a user's historical interactions and a set of candidate items as input, LLMs perform fine-grained relevance estimation to re-rank candidate items, improving recommendation quality beyond traditional methods \cite{llmrank, rec-gpt, lm-as-recsys, chatgptgoodrecommender}.


To provide richer information to the LLM, recent efforts have focused on \textit{user profiling}, which generates representative and descriptive textual profiles of users. 
As raw interaction data is inherently noisy and only indirectly reveals user preferences, merely injecting such data into prompts often yields suboptimal results \cite{llm-trsr}.
Instead, user profiling consolidates historical signals into structured, high-level representations that more faithfully capture underlying preferences.
Specifically, LLMs leverage their reasoning capabilities to infer user preferences from past interactions \cite{palr, llmrec, sprint}, reviews \cite{exp3rt, langptune}, and demographic features \cite{llm4rerank, kar}. 
These profiles articulate latent preferences not explicitly observable from raw interactions alone, such as specific likes and dislikes \cite{exp3rt}, behavioral habits \cite{llm-trsr}, and long- and short-term interests \cite{deeper}.
Prior work shows that integrating these user profiles into input prompts facilitates a deeper understanding of users and improves LLM-based reranking~\cite{languagebaseduserprofilesrecommendation, lettingo}.



Despite progress on the user-side, comparatively little effort has been devoted to \textit{item-side profiling}.
In real-world applications, each item is associated with numerous attributes, from tens to even hundreds, spanning diverse aspects, from functionality (e.g., battery size, WiFi version) to aesthetics (e.g., color, material).
As evidenced by prior work on feature-based recommendation \cite{xdeepfm, wide-deep, AdaFS}, a deep understanding of item content and fine-grained user-item feature interactions is critical.
However, LLMs struggle to utilize such item-side information due to three key challenges:

\vspace{0.05cm}
\noindent\challenge{1} \textbf{Vast unstructured data overwhelm LLM context window}:
Item-side information is vast in scale and often inconsistently formatted across providers, appearing in diverse and non-standardized forms (e.g., bullet lists, tables, or free-form text).
Moreover, many features are spread across long textual descriptions rather than being clearly structured.
LLMs operate under strict input context limits, and struggle to process information within long contexts \cite{lostinthemiddle}.
Therefore, simply including all item data not only wastes the limited context budget but can also lead to suboptimal results.

\noindent\challenge{2} \textbf{Important features differ across items}:
The importance of item features is highly item-dependent.
For example, think of the refresh rate of monitors.
Most monitors share common specifications such as a 60Hz refresh rate, and therefore it is not a notable feature for most items in the category.
However, it becomes a distinguishing factor for a specific subset, like gaming monitors at 360Hz.
Therefore, prompts that include the same set of attributes for all items can obscure truly discriminative attributes with generic ones, making it harder for LLMs to judge fine-grained relevance.


\noindent\challenge{3} \textbf{Even for a single item, important features vary by user}:
Even among the salient features of a single item, different users may find different features appealing.
This user-dependent variation arises from differences in their preferences and usage contexts.
For example, consider Apple’s AirPods Max: 
users with strong brand loyalty to Apple may value its seamless integration within the Apple ecosystem, whereas some may prioritize functional aspects such as high-quality noise canceling.
Given that feature importance varies across users, ignoring such variation can lead to important features for a given user being underrepresented, making it harder for LLMs to accurately assess personalized item relevance.


Still, existing LLM-based reranking approaches only partially address these challenges.
Many methods rely on item titles \cite{msl, deeper, weaverec} or a small set of manually selected attributes \cite{P5, llm4rerank, palr, upr, netflix-llm-reasoning, lettingo, llm-trsr, recsaver}.
They not only provide an incomplete understanding of items but also fail to scale to vast metadata, as they merely inject the given features, leaving \challenge{1} unresolved.
Another line of work constructs item-specific profile
by summarizing item information into concise forms, such as keywords \cite{thinkrec}, motivational traits \cite{m-llm3rec}, and review-based liked/disliked points \cite{exp3rt, reason4rec}.
While scalable, these approaches still rely on a fixed summarization scheme (e.g., extracting ten keywords of each item~\cite{thinkrec}) applied to a small set of manually selected attributes.
This limits their ability to identify item-specific salient features from vast metadata and reviews, leaving \challenge{2} insufficiently addressed.
More importantly, as they generate a \textit{static} description of an item regardless of the user, they fail to capture user-dependent variations in feature salience, leaving \challenge{3} underexplored.

As a solution, we propose \textbf{\proposed}, a user  \textbf{C}ontext-\textbf{A}ware \textbf{I}tem p\textbf{RO}filing framework for LLM-based reranking.
\proposed first transforms large-scale, heterogeneous item metadata into a unified structured representation by extracting domain-specific \textit{objective features} and multi-faceted \textit{subjective traits} from metadata and reviews.
It then builds a lightweight item profiler that, given the current user's information, selects the objective and subjective information relevant to that user's decision-making context.
In particular, rather than relying on fixed heuristics, the profiler learns to quantify the importance of features from collaborative signals, allowing it to identify user-relevant information from vast metadata in an efficient manner.
In this way, \proposed avoids relying on fixed manually selected attributes or static item summaries, while enabling scalable and user-adaptive item profiling for LLM reranking, resolving all the aforementioned challenges.
Furthermore, \proposed incorporates a dedicated refinement process to mitigate potential deficiencies in LLM-inferred information and further improve the quality of item profiling.
Our main contributions are as follows:
\vspace{-\topsep}
\begin{itemize}[leftmargin=0.9em, noitemsep]
    \item We identify an underexplored challenge in LLM-based reranking: effectively leveraging vast and heterogeneous item metadata while capturing item- and user-dependent feature salience.
    \item We propose \proposed, a novel item profiling framework that transforms raw metadata into objective features and subjective traits, refines them, and selectively constructs user-specific item profiles through lightweight selectors.
    \item Extensive experiments show that our framework consistently improves LLM-based reranking, highlighting the value of scalable and user-adaptive item profiling.
\end{itemize}

\vspace{-0.4cm}
\section{Related Work}
\vspace{-0.05cm}
\label{sec:relatedwork}

\smallsection{LLM-based recommendation.}
Equipped with vast world knowledge and reasoning capability, LLMs are widely adopted in recommendation.
For example, LLMRank~\cite{llmrank} shows that LLMs, even when given only item titles, can effectively rank candidate items by leveraging their parametric knowledge.
To exploit LLMs as recommenders, many methods~\cite{tallrec, P5, collm, thinkrec} fine-tune LLM for specific downstream tasks.
Although fine-tuning can align an LLM's knowledge and the reasoning process with a target recommendation task, it requires substantial computational resources and risks compromising the model's generalization ability~\cite{recsys-llmera}.

The prompt augmentation paradigm instead incorporates auxiliary information into the input prompts, thereby preserving the generalization ability of pretrained LLMs.
To enrich prompts, prior studies have explored diverse sources of auxiliary information, with each framework adopting different forms of augmentation.
For the user-side, prior work has incorporated demographic information~\cite{llm4rerank}, user histories~\cite{llm-trsr, palr}, and reviews~\cite{upr, exp3rt}.
For the item-side, prior studies have explored selected attribute fields, such as category~\cite{llm-trsr}, brand and category~\cite{recsaver}, and genre~\cite{lettingo}, as well as graph-derived knowledge~\cite{kg-llm-rec}.
This paradigm enables pretrained LLMs to generalize across diverse recommendation domains without costly fine-tuning, and our work also follows this direction.

\smallsection{User and item profiling in recommendation.}
To provide rich information in prompts, textual compression of useful information, referred to as profiling, has been widely studied on the user side~\cite{upr, llm-trsr, palr, deeper, lettingo}.
Early studies~\cite{llm-trsr, palr} summarize user histories into a static profile, while later work explores more adaptive forms of profiling:
~\cite{upr} allows users to update their profiles explicitly,
~\cite{deeper} updates profiles to reflect evolving preferences, 
and~\cite{lettingo} studies profile formats tailored to downstream tasks.
These efforts help LLMs better understand users and improve~recommendation~accuracy.

By contrast, item-side profiling in LLM-based recommendation has received comparatively less attention.
Most previous studies resort to a fixed set of item attributes~\cite{palr, llm-trsr, lettingo},
or construct item profiles using simple keywords summarized from descriptions~\cite{thinkrec, agentic-tagger} or reviews~\cite{reason4rec}.
Among recent efforts to advance item profiling, the most closely related are EXP3RT~\cite{exp3rt} and M-LLM\textsuperscript{3}Rec~\cite{m-llm3rec}.
In addition to a few selected attributes, EXP3RT extracts preference information from reviews and aggregates them per item, to capture subjective perspectives.
Similarly, M-LLM\textsuperscript{3}Rec uses phrases containing functional purposes of an item as item profiles.

Although these methods are state-of-the-art item profiling methods that effectively capture valuable item information, they have limited capability in identifying salient features of each item from vast metadata, and they use a static item profile for all users, making LLMs struggle to judge personalized item relevance.
In this work, we propose an item profiling strategy that effectively exploits vast item metadata and selectively includes item information relevant to recommendation contexts.



\vspace{-0.2cm}
\section{Problem Formulation}
\vspace{-0.05cm}
\label{sec:preliminary}



\smallsection{Notations.}
Let $\mathcal{U}$ and $\mathcal{I}$ denote the set of users and items in a recommendation dataset, respectively. 
Each user $u \in \mathcal{U}$ has an interaction history $\mathcal{H}_u = \{ r_1, \ldots, r_{|\mathcal{H}_u|} \}$.
Each item $i \in \mathcal{I}$ is associated with metadata $\mathcal{M}_i$ and a set of user reviews $\mathcal{R}_i$.

In many real-world applications, $\mathcal{M}_i$ consists of a large collection of unprocessed information collected from various sources.
It is vast in scale, often containing tens to hundreds of attributes spanning diverse aspects~\cite{kuairand-dataset, yahoo-news-dataset}.
Moreover, item metadata is often inconsistently formatted across providers, appearing in non-standardized forms such as bullet lists, tables, or free-form text.
These attributes span multiple types, including numerical (e.g., price, ratings), categorical (e.g., brand, color), and ordinal values (e.g., size levels, quality tiers).
Furthermore, many important signals are not explicitly structured but are buried in long textual descriptions (Figure~\ref{fig:concept}).



\smallsection{Recommendation with LLM-based reranking.}
We adopt the standard two-stage recommendation \cite{mrs-reranking, ali-rerank}, consisting of candidate generation followed by reranking.
Let $u$ be a user with history $\mathcal{H}_u$, and $\mathcal{C}_u$ be the set of candidate items for $u$ retrieved from the full item set using a lightweight model.
LLMs rerank items in $\mathcal{C}_{u}$ based on their relevance to user $u$, generating a reranked list~$\mathcal{C}_{u}^*$:
\begin{equation}
    \mathcal{C}_u^* = LLM([P_i \mid i \in \mathcal{H}_u], \{P_i \mid i \in \mathcal{C}_u\}, P_u; p_{\text{rerank}}),
\end{equation}
where $p_{\text{rerank}}$ is a prompt for the reranking task.
$P_u$ and $P_i$ denote textual profiles of user and item, respectively.
As aforementioned, while many efforts have focused on $P_u$, constructing $P_i$ that fully leverages $\mathcal{M}_i$ remains underexplored.

\smallsection{Problem definition.}
Our goal is to develop an item profiling function $\phi$ that constructs $P_{i|u} = \phi(\mathcal{M}_i, u)$, where $P_{i|u}$ is a compact yet informative profile of item $i$ \textit{specialized} for user $u$.
We aim to produce profiles that address the aforementioned challenges:  \challenge{1} being structured for limited context window, \challenge{2} capturing item-specific salient features, and \challenge{3} adapting to user context.


\vspace{-0.05cm}
\section{Proposed framework: \proposed}
\label{sec:method}


\begin{figure*}[t]
    \centering    
    \includegraphics[width=1.02\textwidth]{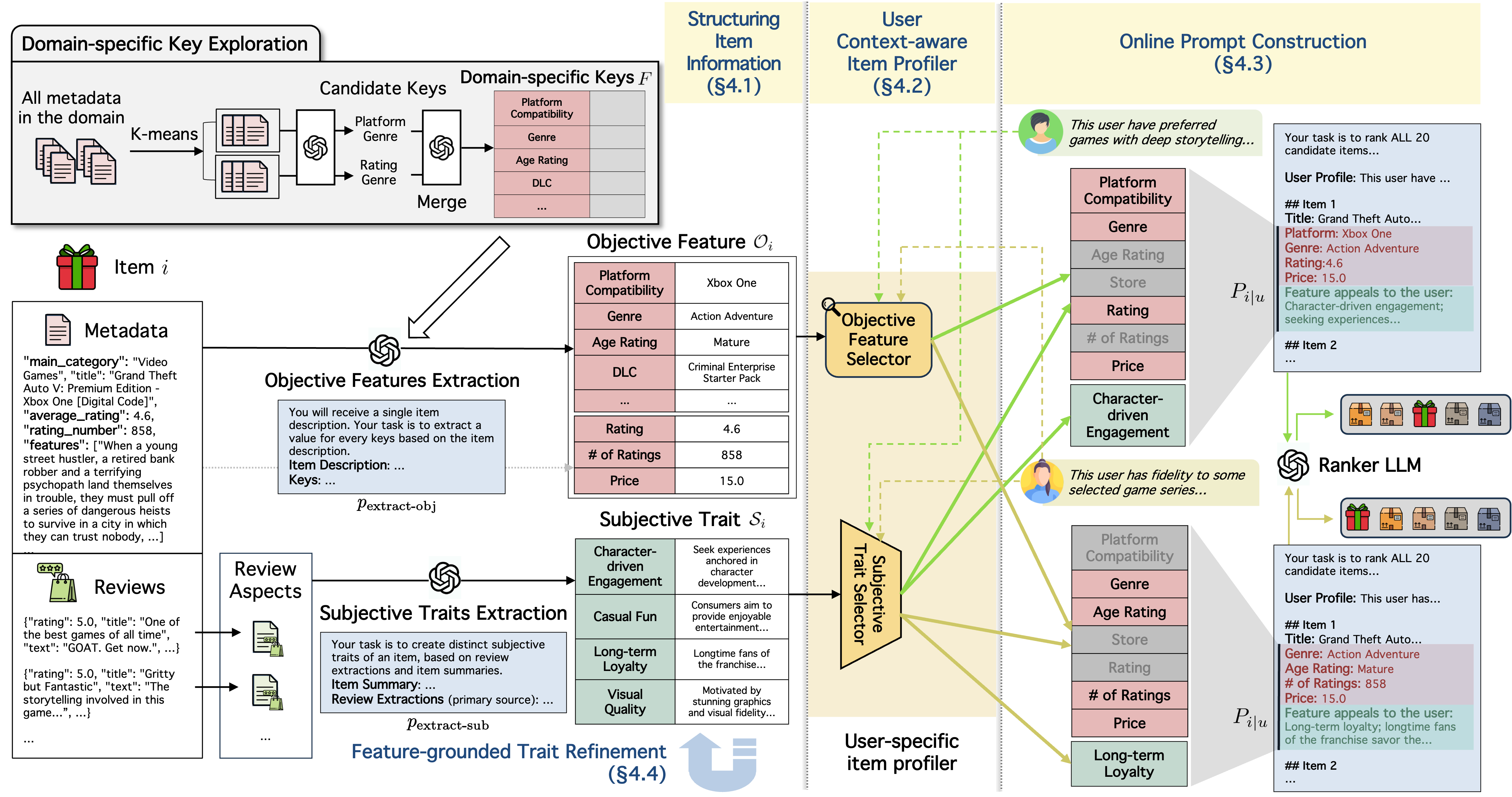}
    \caption{Overview of the \proposed framework.
    }
    \label{fig:method}
    \vspace{-0.5cm}
\end{figure*}


We propose \proposed, which constructs structured and context-specific item profiles for LLM-based reranking. 
We first organize vast item-side knowledge into structured representations (\cref{sec:extraction}), and then build a lightweight profiler that selects context-relevant information (\cref{sec:item-profiler}).
Using this profiler, \proposed generates context-specific item profiles online with relatively negligible latency (\cref{online-prompt-construction}).
We further introduce an optional refinement stage to further enhance the recommendation quality (\cref{sec:refinement}).

\subsection{Structuring Item Information}
\label{sec:extraction}
Raw item metadata $\{\mathcal{M}_{i}\}_{i \in \mathcal{I}}$ comprise large-scale, heterogeneous information from diverse sources, and item reviews $\{\mathcal{R}_i\}_{i \in \mathcal{I}}$, written by diverse users, vary widely in writing styles and expressions, making both highly unstructured and difficult to directly utilize. 
As a first step, we organize item knowledge into a structured form, represented as a dictionary, facilitating systematic usage and compatibility with LLM prompting in the later stages.

Specifically, we consider item knowledge to be two-fold: \textit{objective} and \textit{subjective}.
Objective knowledge refers to factual information that is static across users and is primarily derived from item metadata, such as \verb|{Weight: 3kg}|.
In contrast, subjective knowledge captures aspects that can be perceived differently across users and is mainly inferred from user reviews, such as \verb|{Brand loyalty|: \verb|The brand has strong heritage...}|.
To represent these complementary aspects, we construct two types of records for each item, objective features and subjective traits.


\vspace{-0.1cm}
\subsubsection{\textbf{Domain-specific Key Exploration}}
To extract objective features from metadata, we first define feature fields that serve as keys in the structured dictionary.
These keys should be tailored to the domain, capturing essential aspects in a consistent manner.
Here, the challenges are two-fold:
(i) feature fields in raw metadata are often inconsistent, with semantically identical attributes expressed under different names (e.g., \verb|DLC| vs. \verb|Additional Content|);
(ii) important attributes are often not explicitly provided but are instead embedded in textual descriptions.
For example, in the video game domain, features such as \verb|co-op support| are often mentioned only in text descriptions.
Therefore, we first identify a set of representative keys for the domain, referred to as \textit{domain-specific keys}, and then organize scattered item information based on them.

A straightforward approach is to prompt an LLM with the domain name and directly generate keys.
However, such an approach may fail to cover all subcategories within the domain and is prone to hallucination.
Instead, we adopt a data-driven strategy that lets the LLM extract keys grounded in the actual metadata.
To ensure balanced coverage of diverse subcategories within a domain, we first identify clusters of similar items, generate candidate keys for each cluster, and then merge them into a unified set.

We apply $k$-means clustering to the embeddings of all item metadata.\footnote{We linearize the raw metadata of each item and encode it using BGE-M3 \cite{bge-m3}.}
For each cluster $\mathcal{I}_j$, we randomly sample a small set of items $\tilde{\mathcal{I}}_j \subseteq \mathcal{I}_j$ and let LLM identify representative keys for each cluster:
\begin{equation}
	\mathcal{F}_{j} = LLM(\{ \mathcal{M}_i \mid i \in \tilde{\mathcal{I}}_j \}; p_{\text{explore-keys}}), \qquad j = 1, 2, \ldots, k_{\text{items}},
\end{equation}
where $p_{\text{explore-keys}}$ is the prompt for key exploration.
This yields $k_{\text{items}}$ sets of candidate keys, each reflecting cluster-specific characteristics.
We set $k_{\text{items}} = 20$.

%

We then reassess and consolidate the candidate keys to build a unified set.
Using its reasoning capability, the LLM evaluates whether each key is relevant to user decision-making in the domain, while merging semantically similar keys across clusters (e.g., \verb|Water-proof| vs. \verb|Water-resistance|), as:
\begin{equation}
    \label{eq:reassess}
    \mathcal{F} = LLM(\{ \mathcal{F}_j \mid j = 1, 2, \ldots, k_{\text{items}}\}; p_{\text{reassess-\&-merge}}),
\end{equation}
where $p_{\text{reassess-\&-merge}}$ is the corresponding prompt.
Note that we do not apply frequency-based filtering, so as to preserve sparse but informative keys and ensure broad coverage within the domain.
The resulting set $\mathcal{F}$ defines the final domain-specific keys that comprehensively cover item characteristics within the domain.
In our experiments, $|\mathcal{F}|$ ranges from 30 to 50 depending on the dataset.





\vspace{-0.1cm}
\subsubsection{\textbf{Objective Item Features Extraction}}
Based on the domain-specific keys $\mathcal{F}$, we organize the metadata of all items under a unified schema.
Specifically, for each item $i$, we prompt the LLM to extract the value of each key in $\mathcal{F}$, producing a structured dictionary of objective item features:
\vspace{-0.05cm}
\begin{equation}
	\mathcal{O}_{i} = LLM(\mathcal{M}_i, \mathcal{F}; p_{\text{extract-obj}}), \qquad i \in \mathcal{I},
\end{equation}
where $\mathcal{O}_i = \{ (f: v_f) \mid f \in \mathcal{F} \}$ denotes the resulting feature dictionary.
$\mathcal{O}_i$ includes not only attributes that are already structured in raw metadata (e.g., price), but also diverse attributes embedded in textual descriptions.
To prevent hallucination, we design the reasoning process such that the LLM first \textit{verifies} whether supporting evidence for each key can be found in the metadata, and \textit{assigns} a value only when such evidence is found; otherwise, it outputs~\verb|null|.


\vspace{-0.15cm}
\subsubsection{\textbf{Subjective Item Traits Extraction}}
Subjective traits capture aspects of an item that may be perceived differently across users.
We use user reviews as the primary source of such traits, as they reflect subjective experiences, preferences, and reasons for purchase.
Inspired by prior work that extracts aspect-level signals from reviews (e.g., keywords~\cite{hadsf}, liked/disliked points~\cite{exp3rt}), we first leverage the LLM to convert reviews written in varied styles into a unified schema.
Specifically, for each review $r$, the LLM produces a structured representation $a_r$ consisting of four aspects: \textit{purpose of purchase}, \textit{usage context}, \textit{degree of satisfaction}, and \textit{category preference}. 
As in objective feature extraction, if a particular aspect cannot be identified from the review, the LLM outputs \verb|null|.

Now, item traits can be constructed by summarizing the extracted aspects for item $i$'s reviews: $\{ a_r \mid r \in \mathcal{R}_{i} \}$.
However, we argue that a na\"ive single summary is insufficient to capture the diverse subjective perspectives, as users may value different facets, such as functionalities, aesthetic design, or brand affinity.
We therefore model subjective traits as a multi-faceted set, where each trait captures a distinct perspective commonly reflected in user reviews.
To this end, we design the reasoning process so that the LLM infers multiple traits by reasoning over the extracted aspects, while grounding them in the item's objective features:
\vspace{-0.15cm}\begin{equation}
	\mathcal{S}_{i} = LLM(\{ a_r \mid r \in R_{i} \}, \mathcal{O}_i \,;\, p_{\text{extract-sub}}), \qquad i \in \mathcal{I}
\end{equation}
The prompt $p_{\text{extract-sub}}$ constrains the number of generated traits $|\mathcal{S}_i|$ to 3--7 to avoid collapsing into a single trait or generating redundant ones.
In addition, for each trait, we identify a small set of \textit{supporting features} from $\mathcal{O}_i$ that directly support the trait. 
For example, a trait related to `Vivid Color' can be supported by the feature \verb|{color: red}|.
In practice, each trait is associated with about two objective features on average.

The resulting $\mathcal{S}_{i}$ is a dictionary, i.e., $\mathcal{S}_{i} = \{ t_{i_m}: (d_{i_m} , \mathcal{O}_{i_m} )\}_{m=1}^{|\mathcal{S}_i|}$, where $t_{i_m}$ denotes the trait title, $d_{i_m}$ its description, and $\mathcal{O}_{i_m}$ its supporting features.
Figure~\ref{fig:method} shows a video game whose subjective traits are multi-faceted: `Character-driven Engagement' reflects experiences anchored in character development, while `Long-term Loyalty' captures the appeal to long-time franchise~fans.



\smallsection{Remarks on potential deviations.}
Despite the explicit grounding on reviews and factual item information, the generated subjective traits may deviate from true user intents.
This is because, unlike objective features obtained by directly identifying corresponding values, subjective traits are extracted through the LLM's reasoning.
To cope with such potential deviations, \proposed also introduces an auxiliary refinement process for subjective traits (Section \ref{sec:refinement}).

\begin{figure*}[t]
    \centering    
    \vspace{-0.2cm}
    \includegraphics[width=1.02\textwidth]{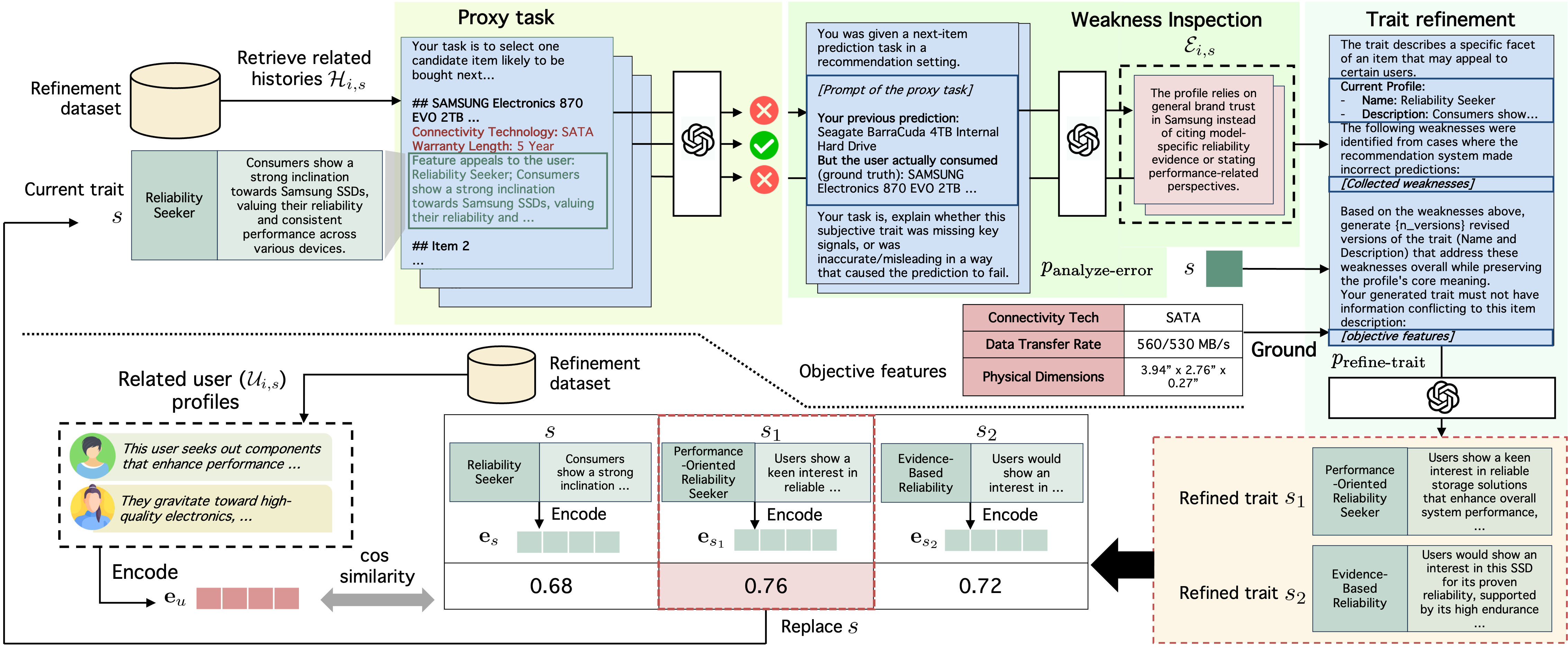}
    \caption{Overview of the feature-grounded trait refinement stage (Section~\ref{sec:refinement}).}
    \label{fig:refine}
    \vspace{-0.5cm}
\end{figure*}


\vspace{-0.1cm}
\subsection{User Context-aware Item Profiler}
\label{sec:item-profiler}
We now obtain structured objective and subjective information for each item.
However, providing all this information to LLMs is suboptimal, as the aspects most relevant to decision-making vary across items and users.
We aim to profile each item using only the information relevant to each user's context and decision-making.
A na\"ive solution is to let LLMs perform this selection at serving time, but this incurs high latency due to additional online~LLM~calls.

To address this, we introduce a lightweight item profiler that generates a user-specific item profile based on the current user's information.
It consists of two submodules for selecting objective features and subjective traits, respectively.
Prepared \textit{offline}, the profiler enables relatively efficient online selection without additional LLM~calls.
In this section, we describe the profiling mechanism used to select objective features and subjective traits, and the final online prompt construction is presented in Section~\ref{online-prompt-construction}.





\vspace{-0.1cm}
\subsubsection{\textbf{User Profile Construction}}
LLM-based rerankers typically take a textual user profile $P_u$ as input (Eq.~1).
Accordingly, we use it as the main source of user information.
As constructing elaborate user profiles is not our main contribution, we largely follow a simple approach from prior work \cite{hadsf, langptune}, which summarizes user reviews for profiling.
Similar to subjective trait extraction, we aggregate review aspects for each user, and prompt the LLM to summarize the user's preferences and interests:
\vspace{-0.05cm}
\begin{equation}
	P_{u} = LLM(\{ a_r \mid r \in \mathcal{R}_{u} \}; p_{\text{extract-user-profile}}), \qquad u \in \mathcal{U},
\end{equation}
where $\mathcal{R}_{u}$ is a set of reviews written by user $u$. 
For the user profiling prompt $p_{\text{extract-user-profile}}$, we follow the prompt used in~\cite{langptune}.



\vspace{-0.1cm}
\subsubsection{\textbf{Sub-module: Objective Feature Selector}}
\label{sec:obj-feat-selector}
For each user~$u$, this sub-module aims to select a small number of features from $\mathcal{O}_i$.
To make such selection effective, it is essential to consider both the user's context and its high-level relationships with item features.
As multiple features may contribute collaboratively rather than independently, it is infeasible to manually define deterministic rules.
We need a mechanism that automatically quantifies which features are most relevant to each user's decision-making.

We draw inspiration from a line of research on quantifying feature importance from collaborative signals~\cite{AdaFS, autofield, mvfs}.
A wisdom from this literature is that, during training, a model can estimate the importance of each feature based on its contribution to prediction.

\vspace{-0.05cm}
\smallsection{A backbone recommender.}
Following this idea, we use a lightweight recommendation model $g_{\text{RS}}(\cdot)$, which predicts the interaction score as $\hat{y}_{ui} = g_{\text{RS}}(\mathbf{e}_{ui})$.\footnote{We use a 2-layer MLP, following prior work~\cite{AdaFS, reaction}.}
Here, $\mathbf{e}_{ui}$ denotes the input representation of the user--item pair $(u,i)$.
By default, $\mathbf{e}_{ui}$ is formed by encoding the user and item information and concatenating the resulting embeddings.
Specifically, we encode each item feature in $\mathcal{O}_i$ as $\mathbf{h}_{f}$ and user profile $P_u$ as $\mathbf{h}_{P_u}$, using encoders corresponding to their modalities.\footnote{Specifically, numerical values are binned, categorical values are embedded, and textual values are encoded with a text embedding model~\cite{bge-m3} followed by a projection layer.}
Also, we treat the user and item IDs as categorical features with embeddings $\mathbf{h}_u$ and $\mathbf{h}_i$, respectively.
Concatenating them yields:
\begin{equation}
    \mathbf{e}_{ui} = [\mathbf{h}_{f_1}; \mathbf{h}_{f_2}; \ldots; \mathbf{h}_{f_n}; \mathbf{h}_{P_u}; \mathbf{h}_u; \mathbf{h}_i] \in \mathbb{R}^{(|\mathcal{F}| + 3) \cdot d}.
\end{equation}

\smallsection{Feature importance learning.}
We introduce a \textit{controller network} $g_w : \mathbb{R}^{(|\mathcal{F}|+3) \cdot d} \rightarrow \mathbb{R}^{|\mathcal{F}|+3}$ which estimates the importance of each feature in $\mathbf{e}_{ui}$, as:
\begin{equation}
\mathbf{w} = \mathrm{softmax}(g_w(\mathbf{e}_{ui})) = [w_{f_1}, w_{f_2}, \ldots, w_{f_n}, w_{P_u}, w_u, w_i],
\end{equation}
where each weight indicates the estimated importance of the corresponding feature.
A weighted input representation $\tilde{\mathbf{e}}_{ui}$ is then constructed by multiplying each importance score with its corresponding embedding:
\begin{equation}
    \tilde{\mathbf{e}}_{ui} = [w_{f_1}\mathbf{h}_{f_1}; w_{f_2}\mathbf{h}_{f_2}; \ldots ;w_{f_n}\mathbf{h}_{f_n}; w_{P_u}\mathbf{h}_{P_u}; w_{u}\mathbf{h}_{u}; w_{i}\mathbf{h}_{i}].
\end{equation}
With the reweighted features, the model makes prediction as: $\hat{y}_{ui} = g_{\text{RS}}(\tilde{\mathbf{e}}_{ui})$, and is optimized using the binary cross-entropy loss:
\vspace{-0.05cm}
\begin{equation}
\mathcal{L}_{\text{BCE}} = - \sum_{(u,i)} \bigl[ y_{ui}\log \hat{y}_{ui} + (1-y_{ui})\log(1-\hat{y}_{ui}) \bigr].
\end{equation}
where $y_{ui}=1$ if $(u,i)$ is observed, and $0$ otherwise.
During training, the recommender and controller are optimized jointly, leading the controller to assign high weights to the objective features most predictive for each user--item pair~\cite{AdaFS}.
This enables automatic quantification of feature importance from collaborative patterns.





\vspace{-0.05cm}
\smallsection{Feature selection.}
For each $(u,i)$, we apply a pooling operation to $\mathbf{w}=\mathrm{softmax}(g_w(\mathbf{e}_{ui}))$ to select the most important features:
\begin{equation}
\mathcal{F}_{u,i} = \mathrm{Pool}(\mathcal{F}, \mathbf{w})
\end{equation}
Following~\cite{AdaFS}, we employ $k$-max pooling, which selects the top-$k$ important features.
In our experiments, we set $k = 4$.
By grounding feature selection in collaboratively learned importance weights, this process can capture complex combinations of features that jointly influence user decisions.




\vspace{-0.05cm}
\subsubsection{\textbf{Sub-module: Subjective Trait Selector}}
For each user~$u$, this sub-module aims to select the subjective trait from $\mathcal{S}_i$ that is most aligned with the user's preferences.
As subjective traits are already expressed as elaborated textual descriptions, we select the most aligned trait based on its semantic similarity to the user profile:
\vspace{-0.15cm}
\begin{equation}
	s_{u,i} = \argmax_{s \in \mathcal{S}_i} \mathrm{cos} (\mathbf{e}_{P_u}, \mathbf{e}_s),
\end{equation}
where $\mathbf{e}_{P_u}$ and $\mathbf{e}_s$ denote the embeddings of the user profile and subjective trait from the text embedding model, respectively.
The selected trait $s_{u,i}$ serves as the subjective aspect of the user-specific profile of item $i$.



\vspace{-0.1cm}
\subsection{Online Prompt Construction}
\label{online-prompt-construction}
Using the proposed mechanisms, we construct the final reranking prompt with item profiles specialized to each user's context.

\setlength{\textfloatsep}{0pt}
\begin{algorithm}
\small
\caption{Online Prompt Construction by \proposed}
\KwIn{
User profile $P_u$, history $H_u$, candidates $C_u$, 
objective features $\{\mathcal{O}_i\}_{i}$, subjective traits $\{\mathcal{S}_i\}_{i}$
}
\KwOut{Final reranking prompt $p_u^{\text{final}}$}

Batch items for parallel processing $B_u \leftarrow H_u \cup C_u$\\

\BlankLine
 \tcp{Objective feature selection}
Construct the input matrix $\mathbf{E}_{u} = [\mathbf{e}_{ui}]_{i\in B_u}$ \hfill (Eq.~7)\\
Compute the importance matrix $\mathbf{W}_{u} = \text{softmax}(g_w(\mathbf{E}_{u}))$ \hfill (Eq.~8)\\
Apply row-wise pooling on $\mathbf{W}_{u}$ to obtain $\{\mathcal{F}_{u,i}\}_{i\in B_u}$ \hfill (Eq.~11)\\

 \tcp{Subjective trait selection}
Construct the trait embedding matrices $\{[\mathbf{E}_s]_{s\in \mathcal{S}_i}\}_{i \in B_u}$\\
Select $s_{u,i} \leftarrow \arg\max_{s \in \mathcal{S}_i} \mathrm{cos}(\mathbf{e}_{P_u}, \mathbf{E}_s)$ for each $i \in B_u$\hfill (Eq.~12)\\
Collect the supporting features $\{\mathcal{F}_{s_{u,i}}\}_{i \in B_u} \leftarrow \{\mathcal{O}_{i_{s_{u,i}}}\}_{i \in B_u}$ \\

\BlankLine
 \tcp{Prompt construction}
Assemble $\{P_{i|u}\}_{i\in B_u}$ with
$P_{i|u} = [\, s_{u,i},\ \mathcal{F}_{u,i} \cup \mathcal{F}_{s_{u,i}} \,]$ \\ 
$p_u^{\text{final}} \leftarrow \mathrm{Prompt}(\{P_{i|u}\}_{i\in H_u},\ \{P_{i|u}\}_{i\in C_u},\ P_u)$

\Return{$p_u^{\text{final}}$}\\
\vspace{-0.05cm}
\end{algorithm}







\vspace{-0.05cm}
\smallsection{Construction process. }
Algorithm 1 details the construction process of the reranking prompt $p_u^{\text{final}}$ of user $u$.
Objective features are selected by $g_w$, implemented as a linear layer followed by pooling (Lines~2--4), while the subjective trait is selected via cosine similarity with the user profile (Lines~5--6).
We additionally include the objective features that directly support the selected subjective trait (Line 7).\footnote{In practice, each item profile contains around five objective features in total.}
The resulting profiles are then assembled into the final reranking prompt, while the remaining instructions follow standard LLM-based reranking.
Each item is profiled according to the current user's context, unlike existing approaches that rely on a single static profile for all users.

Figure~2 shows an example of user-specific prompt construction.
For the first user, the selected subjective trait emphasizes character-driven engagement, and the feature selection module identifies objective features such as platform compatibility and rating as relevant to the user’s decision making based on collaborative patterns.
For the second user, the selected trait instead highlights long-term loyalty, while a different set of objective features, including age rating, is selected.
This illustrates that \proposed adapts both subjective traits and objective features to each user-item context.








\vspace{-0.05cm}
\smallsection{Faithfulness of the profiles.}
CAIRO is equipped with multiple mechanisms to ensure profile faithfulness.
Objective features are extracted only when supported by metadata, with unsupported fields assigned \verb|null|.
Subjective traits are inferred from review-derived aspects and grounded in objective item features, and the final profile includes supporting features for the selected trait.
As a result, all profile information is tightly grounded in the given item information, making the profiles more traceable and less prone to hallucination than free-form item summaries.

\vspace{-0.05cm}
\smallsection{Online efficiency of \proposed. }
A key requirement at this stage is low latency.
To satisfy this, \proposed prepares all item-side information and profiling modules offline, so that the online stage involves only a few lightweight operations.
In particular, all item-side embeddings are precomputed and stored offline, and item profiling is executed in matrix form through parallel GPU computation.
Through this lightweight design, \proposed constructs user-specific item profiles with relatively limited additional overhead in real time.
A detailed efficiency analysis is provided in Section~\ref{sec:comp-cost-anal}.

\vspace{-0.3cm}
\subsection{Feature-grounded Trait Refinement}
\label{sec:refinement}
Although the constructed profiles capture core item aspects, there remains room for improvement, particularly for subjective traits.
Unlike objective features, which are directly evidenced by explicit values, subjective traits are inferred through the LLM’s reasoning process, leaving room for further optimization by aligning them with preference signals from user–item interactions.
To this end, we introduce a trait refinement process that \textit{diagnoses} weaknesses of subjective traits from interaction-based prediction errors and \textit{revises} them accordingly.
The revision is grounded in objective features to ensure faithfulness.
This refinement is conducted offline as an auxiliary step to further improve profile quality.

In alignment with recent studies on prompt optimization \cite{DGDPO, tgd-prompt-opt}, our refinement process consists of three stages: error-case collection, defect diagnosis, and trait refinement.

\vspace{-0.05cm}
\smallsection{Error collection from proxy task.}
To refine each subjective trait,
we first collect cases in which the trait fails to support successful interaction prediction.
To make this process efficient, we adopt simplified next-item prediction as a proxy task.
For each item $i$, we first retrieve histories associated with each subjective trait.
Specifically, from the full interaction histories, we collect subsequences whose last item is $i$ and group them by the corresponding trait.
Let $\mathcal{H}_{i,s}$ denote the set of histories associated with trait $s$ of item $i$.

Given $H \in \mathcal{H}_{i,s}$, we treat $i$ as the ground-truth next item and use the preceding interactions as the user history.
We then ask the LLM to predict the most probable next item from a candidate set that includes item $i$ and $N_{\text{proxy-cand}}$ randomly sampled negative items.
We set $N_{\text{proxy-cand}} = 4$.
The re-ranking prompt is constructed as described in the previous section.
As $i$ is profiled with trait $s$, an error case where the LLM fails to select $i$ provides evidence that the current trait may not sufficiently support preference prediction.

\smallsection{Defect diagnosis.}
From the proxy task, we obtain error cases in which the LLM fails to predict the ground-truth next item.
Let $\mathcal{E}_{i,s} = \{ H_1, H_2, \ldots, H_{|\mathcal{E}_{i,s}|} \}$ denote the set of error cases from $\mathcal{H}_{i, s}$.
We then instruct the LLM to analyze each error case and identify the deficiency of the current trait $s$.
Formally, we obtain the diagnosis $D_k$ for each error case $H_k$ as follows:
\begin{equation}
	D_k = LLM(H_k; p_{\text{analyze-error}}), \quad k = 1, 2, \ldots, |\mathcal{E}_{i,s}|.
\end{equation}
The resulting diagnoses $\{D_k\}_k$ are collectively leveraged to capture failure patterns that consistently appear across multiple error cases.
Figure~3 illustrates an example of this diagnosis step, where the LLM identifies that the current trait overemphasizes general brand reliability while missing performance-related aspects.

\smallsection{Objective feature-grounded trait refinement.}
Using the collected diagnoses, we refine the subjective trait by addressing recurring deficiencies.
To prevent hallucination, we ground the reasoning process in the objective features of each item.
Specifically, the LLM revises the trait by jointly considering the original trait, the diagnosis set, and the objective features.

Following the common practice of sampling multiple LLM outputs and selecting the most suitable one to improve generation quality~\cite{math-shepherd, verify-step}, we generate $N_{\text{cand}}$ candidate revisions:
\begin{equation}
	s'_q = LLM(s, \{D_k\}_{k=1}^{|\mathcal{E}_{i,s}|}, \mathcal{O}_i; p_{\text{refine-trait}}), \quad
    q = 1, 2, \ldots, N_{\text{cand}}.
\end{equation}
We set $N_{\text{cand}} = 2$.
Finally, among the original and generated traits, we select the trait most aligned with the users associated with $\mathcal{H}_{i,s}$.
Alignment is measured by the cosine similarity between each trait embedding and the average embedding of the corresponding user profiles, and the trait with the highest similarity is selected as the refined trait.
Figure~3 shows that the refined trait better captures performance-related aspects aligned with user preferences.
Overall, this refinement process directly reflects prediction failures through diagnosis-based revision, while objective feature grounding preserves faithfulness of profiles.




\vspace{-0.1cm}
\section{Experiments}

\begin{table*}[h]
\centering
\caption{Overall performance comparison (*: $p$-value < .05). For feature-enhanced LLM-based reranking frameworks (+Feat), metrics colored  \textcolor{red}{red} denote under-performance over LLMRank, which uses only item titles.}
\label{tab:main}
{\small
\renewcommand{\arraystretch}{0.95}
\resizebox{\textwidth}{!}{%
\begin{tabular}{@{}ccccccccccccc@{}}
\toprule
\multirow{2}{*}{\textbf{Methods}} & \multicolumn{4}{c} {\textbf{Video Games}} & \multicolumn{4}{c}{\textbf{Sports and Outdoors}} & \multicolumn{4}{c}{\textbf{Electronics}} \\ \cmidrule(l){2-5} \cmidrule(l){6-9} \cmidrule(l){10-13} 
                                  & \footnotesize nDCG@5   & \footnotesize HR@5     & \footnotesize nDCG@10  & \footnotesize HR@10   & \footnotesize nDCG@5     & \footnotesize HR@5       & \footnotesize nDCG@10    & \footnotesize HR@10     & \footnotesize nDCG@5  & \footnotesize HR@5   & \footnotesize nDCG@10  & \footnotesize HR@10  \\ \midrule 

BPR                               & 0.0698   & 0.1211   & 0.1252   & 0.2961  & 0.0793     & 0.1305     & 0.1303     & 0.3082    & 0.0540  & 0.0960 & 0.1086   & 0.2619 \\
SASRec                            & 0.1599   & 0.2205   & 0.1952   & 0.3349  & 0.1658     & 0.2203     & 0.1968     & 0.3172    & 0.1399  & 0.1835 & 0.1650   & 0.2868 \\
BERT4Rec                          & 0.1488   & 0.2123   & 0.1842   & 0.3226  & 0.1247     & 0.1825     & 0.1613     & 0.2967    & 0.0938  & 0.1372 & 0.1215   & 0.2237 \\ \midrule
xDeepFM                           & 0.0932   & 0.1679   & 0.1621   & 0.3834   & 0.0897   & 0.1637   & 0.1638   & 0.3959   & 0.0879   & 0.1491   & 0.1497   & 0.3444 \\
AdaFS                             & 0.1401   & 0.2140   & 0.2030   & 0.4119   & 0.1247   & 0.2109   & 0.1937   & 0.4273   & 0.1029   & 0.1752   & 0.1762   & 0.4054  \\ 
REACTION                          & 0.1404   & 0.2293   & 0.2005   & 0.4185   & 0.1271   & 0.2240   & 0.2035   & 0.4441   & 0.1055   & 0.1901   & 0.1807   & 0.4257  \\ \midrule
LLMRank                           & 0.1626   & 0.2411   & 0.2180   & 0.4154  & 0.1652     & 0.2434     & 0.2131     & 0.3934    & 0.1214  & 0.1779 & 0.1607   & 0.3012 \\
LLMRank \textbf{(+Feat)}                       &  \textcolor{red}{0.1122} &  \textcolor{red}{0.1728} &  \textcolor{red}{0.1527} &  \textcolor{red}{0.3026} &  \textcolor{red}{0.1114} &  \textcolor{red}{0.1725} &  \textcolor{red}{0.1513} &  \textcolor{red}{0.2978} &   \textcolor{red}{0.1070} &  \textcolor{red}{0.1631} &  \textcolor{red}{0.1470} &  \textcolor{red}{0.2885}\\
EXP3RT                            & 0.1644   & 0.2414   & 0.2172   & 0.4073  & 0.1796     & 0.2630     & 0.2453     & 0.4690    & 0.1604  & 0.2276 & 0.2152   & 0.4001 \\
EXP3RT \textbf{(+Feat)}                   &  \textcolor{red}{0.1303} &  \textcolor{red}{0.2052} &  \textcolor{red}{0.1851} &  \textcolor{red}{0.3775} &  \textcolor{red}{0.1420} &  \textcolor{red}{0.2134} &  \textcolor{red}{0.2017} &  {0.4011} &  \textcolor{red}{0.1109} &  \textcolor{red}{0.1713} &  {0.1656} &  {0.3444} \\
M-LLM3Rec                         & 0.1674   & 0.2578   & 0.2290   & 0.4505  & 0.1734     & 0.2667     & 0.2409     & 0.4777   & 0.1581  & 0.2422 & 0.2181   & 0.4308 \\ 
M-LLM3Rec \textbf{(+Feat)}                 &  \textcolor{red}{0.1519} &  \textcolor{red}{0.2373} &  \textcolor{red}{0.2119} &  {0.4202} &  {0.1676} &  {0.2592} &  {0.2309} &  {0.4574} &  {0.1472} &  {0.2304} &  {0.2069} &  {0.4173} \\ \midrule

\textbf{\proposed} & \textbf{\makecell{0.1839* \\ \footnotesize(+9.86\%) }}   & \textbf{\makecell{0.2761* \\ \footnotesize(+7.10\%) }}   & \textbf{\makecell{0.2435* \\ \footnotesize(+6.33\%) }}   & \textbf{\makecell{0.4605* \\ \footnotesize(+2.22\%) }}  & \textbf{\makecell{0.1852* \\ \footnotesize(+3.12\%) }}    & \textbf{\makecell{0.2793* \\ \footnotesize(+4.72\%) }} & \textbf{\makecell{0.2512* \\ \footnotesize(+2.41\%) }} & \textbf{\makecell{0.4865* \\ \footnotesize(+1.84\%) }}  & \textbf{\makecell{0.1741* \\ \footnotesize(+9.98\%) }}  & \textbf{\makecell{0.2624* \\ \footnotesize(+9.06\%) }}  & \textbf{\makecell{0.2405* \\ \footnotesize(+10.47\%) }}   & \textbf{\makecell{0.4706* \\ \footnotesize(+9.09\%) }} \\ 
\textbf{\proposed + Refine} & \textbf{\makecell{0.1895* \\ \footnotesize(+12.84\%) }}  & \textbf{\makecell{0.2827* \\ \footnotesize(+10.40\%) }}   & \textbf{\makecell{0.2467* \\ \footnotesize(+7.51\%) }}   & \textbf{\makecell{0.4642* \\ \footnotesize(+3.15\%) }}  & \textbf{\makecell{0.1981* \\ \footnotesize(+10.30\%) }} & \textbf{\makecell{0.2877* \\ \footnotesize(+7.87\%) }}& \textbf{\makecell{0.2643* \\ \footnotesize(+7.75\%) }} & \textbf{\makecell{0.4959* \\ \footnotesize(+3.81\%) }} & \textbf{\makecell{0.1866* \\ \footnotesize(+17.88\%) }} & \textbf{\makecell{0.2693* \\ \footnotesize(+11.93\%) }} & \textbf{\makecell{0.2540* \\ \footnotesize(+16.67\%) }} & \textbf{\makecell{0.4804* \\ \footnotesize(+11.36\%) }} \\
\bottomrule
\end{tabular}}
}
\vspace{-0.4cm}
\end{table*}


\subsection{Experimental Setup}
\label{sec:experimentsetup}

The key instructions of each prompt are presented in the main text and figures, and the \textbf{full prompts} are provided in the code.\footnote{\repolink}

\vspace{-0.05cm}
\subsubsection{\textbf{Datasets}} 
We conduct experiments on Amazon datasets~\cite{amazon-review-dataset-23}, which, to our knowledge, provide the largest sources of vast and heterogeneous item metadata with user reviews, which allows meaningful analysis on item-side information.
We select three domains with distinct characteristics: Video Games, Sports and Outdoors, and Electronics.
We closely follow the preprocessing of prior studies~\cite{exp3rt, m-llm3rec, llamaRec, tallrec}:
considering the significant cost of generating profiles with competitive commercial LLMs, we sample 200K--300K interactions from each dataset and apply 5-core filtering on both users and items.
To enable meaningful analysis of item-side information, we filter out items with missing titles or metadata.

For item metadata, we use both common and optional features. 
The common features consist of 11 fields observed across all items and datasets, while optional features vary across items and may appear under different provider-specific field names. 
Among the common features, the \verb|features| and \verb|description| fields are written as unstructured sentences.
Detailed dataset statistics and the used metadata fields are reported in Table \ref{tab:dataset-summary}.



\vspace{-0.15cm}
\subsubsection{\textbf{Baselines}}
We consider a variety of baseline methods.

\smallsection{Traditional ranking models.} We compare against three methods:
    \begin{itemize}[label=$\bullet$, leftmargin=1em, nosep]
        \item \textbf{BPR}~\cite{bpr} is an embedding-based model with pairwise ranking.
        \item \textbf{SASRec}~\cite{sasrec} and \textbf{BERT4Rec}~\cite{bert4rec} are transformer-based models with unidirectional/bidirectional attention, respectively.
    \end{itemize}

   

\smallsection{Feature-based models.} We compare against competitive methods that learn feature importance from vast metadata and incorporate fine-grained user--item feature interactions into recommendation:
\begin{itemize}[label=$\bullet$, leftmargin=1em, nosep]
   \item \textbf{xDeepFM} \cite{xdeepfm} explicitly models high-order feature interactions via a cross network with the cross-product operation.
   \item  \textbf{AdaFS} \cite{AdaFS} learns interaction-specific feature importance and selects a small set of important features for prediction.
   \item  \textbf{REACTION} \cite{reaction} is a state-of-the-art method that jointly considers feature redundancy and parameter efficiency by formulating feature selection from an information-theoretic perspective.
\end{itemize}
    

\smallsection{LLM-based reranking models.} 
We compare against state-of-the-art LLM-based reranking methods to examine how different profiling strategies affect reranking capability.
All methods use GPT-4o-mini, and the core difference lies in how each item is profiled.

\begin{itemize}[label=$\bullet$, leftmargin=1em, nosep]
    \item \textbf{LLMRank}~\cite{llmrank} is a fundamental LLM-based reranking method that uses only item titles, aiming to leverage the LLM's parametric knowledge for recommendation.
    
    \item \textbf{EXP3RT}~\cite{exp3rt} employs an advanced profiling that extracts liked/disliked points from reviews to construct both user and item profiles, while also incorporating item features as additional signals.
    
    \item \textbf{M-$\text{LLM}^3$Rec} \cite{m-llm3rec} employs a motivation-oriented profiling  to capture user motivations, summarizing item metadata into keyword-level item profiles and constructing user profiles~from~reviews.
\end{itemize}

\noindent
While EXP3RT and M-$\text{LLM}^3$Rec use item metadata, they remain limited to a small set of preprocessed common features (Table~\ref{tab:dataset-summary}), as item profiling is not their primary focus.
To give them more knowledge in terms of the item-side information, we consider \textbf{feature-enhanced variants of the LLM-based methods}, denoted as \textbf{+Feat}.
Here we additionally provide the not-null optional fields in Table~\ref{tab:dataset-summary} not originally used by each method, after linearizing them.


\vspace{-0.05cm}
\smallsection{Agentic RAG profiling.}
Another possible direction for item profiling is to fully exploit LLM reasoning;
LLMs iteratively determine what information is needed for the current decision, retrieve the relevant evidence, and refine the profile over multiple rounds.
\begin{itemize}[label=$\bullet$, leftmargin=1em, nosep]
    \item \textbf{REAP}~\cite{reap} is an agentic RAG framework that iteratively re-plans sub-tasks and retrieves supporting facts. We provide the available metadata fields for each item as the retrieval space and allow the model to select among them for up to 5 rounds.
\end{itemize}
As the RAG-based approach incurs substantial LLM cost, we report a focused evaluation on a subset of users in Section~\ref{sec:exp-rag}.
Note that we exclude methods that use LLMs only in auxiliary roles (e.g., embedding generation~\cite{llmrec, kar, hadsf}) and fine-tune the LLM itself~\cite{collm, thinkrec}, since our goal is enhancing general-purpose LLMs through profiling.


For our proposed method, we consider two variants: \textbf{\proposed} with profiles having original traits from Section~\ref{sec:extraction}, and \textbf{\proposed + Refine} with profiles having refined traits from Section~\ref{sec:refinement}.


\vspace{-0.15cm}
\subsubsection{\textbf{Evaluation Setting}}
Following prior studies~\cite{llm4rerank, P5}, we use each user’s last interaction for testing, the second-to-last interaction for validation, and the remaining for training.
To prevent information leakage, user and item profiles are generated using only the training data.
For reranking, following prior studies on LLM-based reranking~\cite{llmrank, uncertainty-quantification-llm-rec, filling-the-gaps}, we give 20 candidate items, consisting of one ground-truth item and 19 negative samples from BPR-MF, whose order is shuffled to avoid positional bias.
We report Hit Rate (HR@K) and nDCG (nDCG@K) at $K \in \{5, 10\}$.


\vspace{-0.15cm}
\subsubsection{\textbf{Implementation Details}}
We use gpt-4o-mini for all LLM-based methods, including ours.
EXP3RT~\cite{exp3rt} is also implemented with GPT-4o-mini for both profiling and reranking, without distillation to a smaller model for efficiency, to ensure a fair comparison under the same LLM setting.
We set the temperature to 0.2 in reranking stage to make the results more deterministic, following~\cite{llmrank}, but 0.9 for other stages to promote creativity, following~\cite{m-llm3rec}.
For \proposed, we set $k_{\text{items}}=20$, $k=4$, $N_{\text{proxy-cand}}=4$, and $N_{\text{cand}}=2$.
For the text embedding models, we use BGE-M3 \cite{bge-m3}.
For the objective feature selector, the learning rate is chosen from \{5e-3, 2e-3, 1e-3\}, and other hyperparameters are chosen following~\cite{AdaFS}.
All baseline hyperparameter ranges follow their original papers or official implementations.
All feature-based models use both metadata features and user/item IDs, with the same preprocessing scheme as in our objective feature selector.
We conduct the experiment on a server with Intel Xeon Gold 6338 CPUs and four RTX A5000 GPUs.



\begin{table}[t]
\centering
\caption{Statistics and features of the used datasets.}
\label{tab:dataset-summary}
\renewcommand{\arraystretch}{0.9}
\resizebox{\linewidth}{!}{
\begin{tabular}{lccc}
\toprule
 \textbf{Dataset} &  \textbf{Video Games} &  \textbf{Sports and Outdoors} &  \textbf{Electronics}  \\
\midrule
\# Interactions & 263,782 & 232,923 & 223,173 \\
\# Users & 31,937 & 30,734 & 30,403 \\
\# Items & 9,233 & 19,549 & 13,711 \\ 
Sparsity & 99.91\% & 99.96\% & 99.95\% \\ \midrule
\makecell[l]{Common\\features}
& \multicolumn{3}{c}{\footnotesize\makecell{ 
\texttt{main\_category}, \texttt{title}, \texttt{average\_rating}, \texttt{rating\_number}, \texttt{price}, \texttt{store},  \\\texttt{features}, \texttt{description}, \texttt{categories}, \texttt{parent\_asin}, \texttt{bought\_together}
}} \\
\midrule
\makecell[l]{Optional\\features}
& { \footnotesize \makecell[c]{
\texttt{language},\\
\texttt{rated}, \texttt{genre}, \\
\texttt{batteries}, \\
\texttt{power} \texttt{source}, \\
$\cdots$\\
(total 266)
}}
& {\footnotesize \makecell[c]{
\texttt{color}, \texttt{size},\\
\texttt{material}, \\
\texttt{sport} \texttt{type}, \\
\texttt{model} \texttt{year}, $\cdots$\\
(total 964)
}}
& {\footnotesize \makecell[c]{
\texttt{item weight},\\
\texttt{material},\\
\texttt{color}, \texttt{voltage}, \\
\texttt{country of} \\
\texttt{origin}, $\cdots$\\
(total 705)
}} \\ 
\bottomrule
\end{tabular}}
\end{table}

\vspace{-0.2cm}
\subsection{Effect of the Proposed Profiling Strategy}

\subsubsection{\textbf{Overall Performance Comparison}}

Table \ref{tab:main} compares \proposed to the baselines.
We observe that \proposed consistently outperforms across all datasets and metrics.
In particular, the performance gains over LLM-based reranking baselines are most pronounced at $K=5$, evidencing that structured item profiles help LLMs make finer-grained distinctions among top-ranked candidates.
Furthermore, \proposed already outperforms all baselines even without refinement, supporting the effectiveness of the core profiling strategy; the refinement stage further provides consistent additional gains, serving as a complement to the core profiling design.

A key observation is that feature-enhanced variants (+Feat) of existing LLM-based methods degrade across all datasets and metrics, often falling below LLMRank, which uses only item titles.
This supports that na\"ively injecting raw metadata can harm LLM reranking by introducing unstructured and possibly noisy information.
In contrast, \proposed benefits from richer item information by structuring it and selecting user-relevant evidence.
This underscores the importance of \textit{what} information is provided to the LLM, rather than simply \textit{how much} information is provided.




\begin{table}[t!]
\centering
\caption{Comparison of RAG-based method (REAP) and ours on Electronics and Sports and Outdoors datasets. Profiling Time stands for the average time taken to make all candidate item profiles of each user, measured by seconds.}
\label{tab:rag}
\renewcommand{\arraystretch}{0.85}
\resizebox{\linewidth}{!}{
\begin{tabular}{lccccc}
\multicolumn{6}{c}{\textbf{Electronics}} \\ \toprule
\textbf{Method} & \textbf{nDCG@5} & \textbf{HR@5} & \textbf{nDCG@10} & \textbf{HR@10} & \textbf{Profiling Time} \\
\midrule
REAP (5-hop) & 0.1844 & 0.2728 & 0.2460 & 0.4657 & 815.70s \\
REAP (1-hop) & 0.1894 & 0.2729 & 0.2556 & 0.4800 & 346.63s \\
\proposed & 0.1892 & 0.2771 & 0.2673 & 0.5214 & \textbf{0.32s} \\
\bottomrule 
\\
\multicolumn{6}{c}{\textbf{Sports and Outdoors}} \\ \toprule
\textbf{Method} & \textbf{nDCG@5} & \textbf{HR@5} & \textbf{nDCG@10} & \textbf{HR@10} & \textbf{Profiling Time} \\
\midrule
REAP (5-hop) & 0.2092 & 0.3043 & 0.2743 & 0.5100 & 736.94s \\
REAP (1-hop) & 0.2109 & 0.3014 & 0.2788 & 0.5143 & 315.87s \\
\proposed & 0.1986 & 0.3000 & 0.2663 & 0.5143 & \textbf{0.25s} \\
\bottomrule
\end{tabular}
}
\end{table}

\subsubsection{\textbf{Comparison with RAG-based Profiling}}
\label{sec:exp-rag}

To compare \proposed to RAG-based profiling,
we consider two REAP variants: REAP (5-hop), which iteratively retrieves information over up to 5 rounds, and REAP (1-hop), which collapses retrieval into a single step.
These variants require repeated LLM calls to identify important information in each context.
Due to the substantial cost of REAP, we randomly sample 700 users from each test set here. 

Table~\ref{tab:rag} reports performance and average per-user profiling time.
We observe that, while both REAP variants require the prohibitively heavy latency due to multiple LLM callings, \proposed achieves a significant reduction in the average profiling time, while still showing competitive performance to them.
This latency gain is attributable to our LLM-free, offline-prepared profiler that requires only lightweight matrix operations at serving time.

\begin{table}[t]
\centering
\caption{Offline and online efficiency comparison.}
\renewcommand{\arraystretch}{0.85}
\label{tab:efficiency}
\resizebox{\linewidth}{!}{
\begin{tabular}{llcccc}
\toprule
\multirow{3}{*}{\textbf{Dataset}} 
& \multirow{3}{*}{\textbf{Method}} 
& \multicolumn{1}{c}{\textbf{Offline}} 
& \multicolumn{3}{c}{\textbf{Online}} \\ \cmidrule(lr){3-3} \cmidrule(lr){4-6}
& 
& \textbf{\# LLM Calls}
& \textbf{Avg. Token}
& \makecell{\textbf{Profiling}\\ \textbf{Time}}
& \makecell{\textbf{Reranking}\\ \textbf{Time}\textsuperscript{\textdagger}} \\
\midrule

\multirow{2}{*}{\makecell[l]{\textbf{Video}\\\textbf{Games}}}
& EXP3RT & 7.55 & 4584.6 & --    & 2.89s \\
& \proposed   & 7.84 & 4112.4 & 0.21s & 3.03s \\

\midrule

\multirow{2}{*}{\makecell[l]{\textbf{Sports and}\\\textbf{Outdoors}}}
& EXP3RT & 7.55 & 4887.6 & --    & 3.15s \\
& \proposed   & 8.18 & 4954.4 & 0.22s & 3.02s \\

\midrule

\multirow{2}{*}{\textbf{Electronics}}
& EXP3RT & 6.97 & 4337.6 & --    & 3.52s \\
& \proposed   & 7.42 & 4100.0 & 0.21s & 3.20s \\

\bottomrule
\end{tabular}}
\begin{tablenotes}
\footnotesize
    \item \textdagger: Reranking time is sensitive to server status; for reference only. Practically both methods have comparable reranking times.
\end{tablenotes}
\end{table}

\subsubsection{\textbf{Computational Cost Analysis}}
\label{sec:comp-cost-anal}

In Table~\ref{tab:efficiency}, we compare computational cost involved offline and online to EXP3RT, a review-based profiling method like ours.
We measure the average number of LLM calling involved in the offline profiling process, the average number of tokens in the final prompts, and the average latencies involved in online profiling and LLM calling, all per a test user.
To measure the reranking time, we made asynchronous LLM calls with the maximum concurrency 5.
We can observe that \proposed incurs a marginally higher number of offline LLM calls, while online serving latency is comparable to EXP3RT without the online profiling.
Together, these results show that \proposed's consistent performance gains over EXP3RT come at a relatively negligible additional cost.

\subsection{Study of \proposed}
\begin{table}[t]
\centering
\caption{Ablation study on key/trait construction strategies. K- and T- stand for the variants applied to key exploration stage and trait extraction stage, respectively.}
\label{tab:key-comparison}
\renewcommand{\arraystretch}{0.85}
\resizebox{\linewidth}{!}{
\begin{tabular}{llcccc}
\toprule
\textbf{Dataset} & \textbf{Strategy} & \textbf{nDCG@5} & \textbf{HR@5} & \textbf{nDCG@10} & \textbf{HR@10} \\
\midrule
\multirow{4}{*}{\textbf{Video Games}}
& K-Frequency  & 0.1732 & 0.2585 & 0.2330 & 0.4461   \\
& K-MergeMore  & 0.1734 & 0.2617 & 0.2402 & \textbf{0.4711}     \\
& T-Single & 0.1840 & 0.2723 & 0.2366 & 0.4370  \\
& \textbf{\proposed} & \textbf{0.1843} & \textbf{0.2778} & \textbf{0.2438} & 0.4622   \\
\midrule
\multirow{4}{*}{\textbf{Electronics}}
& K-Frequency    & 0.1713 & 0.2560 & 0.2359 & 0.4587 \\
& K-MergeMore       & 0.1735 & 0.2618 & 0.2342 & 0.4529 \\
& T-Single    & 0.1715 & 0.2562 & 0.2372 & 0.4329 \\
& \textbf{\proposed}  &\textbf{0.1757} & \textbf{0.2639} & \textbf{0.2406} & \textbf{0.4678} \\
\bottomrule
\end{tabular}
}
\end{table}
\begin{table}[t]
\centering
\caption{Ablation study on features/trait selection strategies. O- and S- mean the strategies applied to the objective feature selection and the subjective trait selection, respectively.}
\label{tab:objsub-comparison}
\renewcommand{\arraystretch}{0.85}
\resizebox{\linewidth}{!}{
\begin{tabular}{llcccc}
\toprule
\textbf{Dataset} & \textbf{Strategy} & \textbf{nDCG@5} & \textbf{HR@5} & \textbf{nDCG@10} & \textbf{HR@10} \\
\midrule
\multirow{6}{*}{\textbf{Video Games}}
& O-Random    & 0.1799 & 0.2701 & 0.2393 & 0.4503 \\
& O-All       & 0.1782 & 0.2643 & 0.2304 & 0.4290 \\
& O-Sim       & 0.1832 & 0.2679 & 0.2394 & 0.4441 \\
& S-Random    & 0.1797 & 0.2658 & 0.2368 & 0.4448 \\
& S-All       & 0.1744 & 0.2555 & 0.2299 & 0.4302 \\
& \textbf{\proposed} & \textbf{0.1843} & \textbf{0.2778} & \textbf{0.2438} & \textbf{0.4622} \\
\midrule
\multirow{6}{*}{\textbf{Electronics}}
& O-Random    & 0.1782 & 0.2518 & 0.2332 & 0.4292 \\
& O-All       & 0.1601 & 0.2279 & 0.2121 & 0.3918 \\
& O-Sim       & 0.1778 & 0.2528 & 0.2351 & 0.4326 \\
& S-Random    & 0.1785 & 0.2541 & 0.2345 & 0.4298 \\
& S-All       & 0.1674 & 0.2393 & 0.2208 & 0.4070 \\
& \textbf{\proposed} & \textbf{0.1797} & \textbf{0.2571} & \textbf{0.2366} & \textbf{0.4359} \\
\bottomrule
\end{tabular}
}
\vspace{0.2cm}
\end{table}

\subsubsection{\textbf{Ablation Study}} We provide ablation study results of \proposed on Video Games and Electronics datasets.

\smallsection{{Effect of key/trait construction.}}
Table~\ref{tab:key-comparison} ablates our key exploration and trait extraction designs in Section~\ref{sec:extraction}.
For domain-specific key exploration, we compare ours with K-Frequency, which selects the same number of keys from the most frequent optional features for each domain, and K-MergeMore, which modifies $p_{\text{reassess-\&-merge}}$ (Eq.~\ref{eq:reassess}) to ignore sparse features and make a smaller key set.
Both generally perform worse than \proposed, supporting that our key generation strategy allows fine-grained distinctions between diverse types of items and helps LLMs reason about item differences.

For trait extraction, T-Single, which generates only one trait per item, similarly degrades performance.
It evidences that user opinions on an item are multi-faceted and therefore it is critical to provide diverse trait candidates from which the selector can identify the one most aligned with each user's context.

\smallsection{{Effect of feature/trait selection.}}
To observe the effect of our feature/trait selection strategy in Section~\ref{sec:item-profiler}, we compare the selection strategy of \proposed to different selection strategies of objective features and subjective traits:
(1) Random randomly samples a subset of objective features or subjective traits;
(2) All includes all available features or traits in the profile.
For objective feature selection, we also consider Sim, which selects the features by cosine similarity between feature embeddings and the user profile embedding.

Table~\ref{tab:objsub-comparison} reports the results on Video Games and Electronics datasets.
For O-Random, we sample four features, matching our selector.
We first observe that \proposed achieves the best performance in all cases, supporting that each module identifies decision-relevant information for each user-item context.
In contrast, O-All and S-All consistently perform worse, evidencing that indiscriminate feature injection causes distraction of LLM rather than mere redundancy.
We also note that the performance gaps remain moderate, likely because the non-ablated component continues to provide well-constructed signals even when the other is ablated.
Lastly, although O-Sim outperforms random selection, it remains below our strategy, showing that objective feature importance is better captured through collaborative patterns than semantic similarity alone.

\begin{figure}[t!]
  \includegraphics[width=0.95\columnwidth, alt="A chart showing transferability results."]{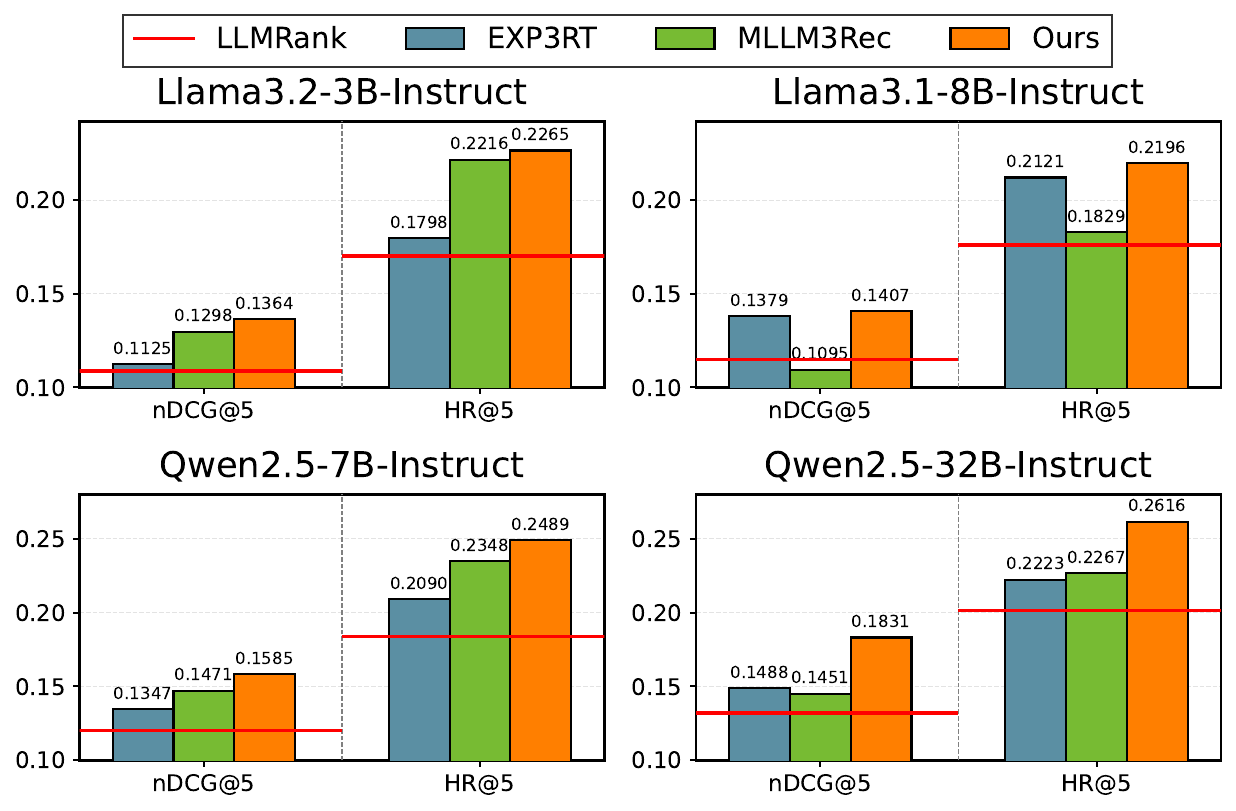}
  \caption{Recommendation performance of diverse LLM-based recommendation methods with profiling on Electronics dataset applied to different families and sizes of LLMs.} 
  \label{fig:transferability}
\end{figure}

\subsubsection{\textbf{Transferability to Other LLMs}}

To examine the transferability of our generated item profiles to diverse LLM families and scales, we conduct experiments using four LLMs spanning two families: Qwen2.5-32B-Instruct, Qwen2.5-7B-Instruct~\cite{qwen2}, Llama3.1-8B-Instruct and Llama3.2-3B-Instruct~\cite{llama3}. 
These models range from 3B to 32B parameters, covering both small and large variants.

The results of the comparison is reported in Figure \ref{fig:transferability}.
\proposed consistently outperforms both baselines across all LLM families and the parameter scales.
This supports that our generated item profiles are not tailored to a specific LLM architecture or scale, but instead hold useful item knowledge that helps diverse LLMs better understand the recommendation context.
Furthermore, \proposed improves even the smallest 3B model, evidencing that well-structured profiles can compensate for limited model capacity.




\begin{table}[t]
\centering
\caption{Case study on Electronics dataset. \textcolor{red}{Red} text can work as signals for functionality, and \textcolor{blue}{blue} text for brand-loyalty.}
\footnotesize
\label{tab:case}
\begin{tabularx}{\linewidth}{L}
\toprule
\textbf{Item}: B0B6WTFTG9 (Google Pixel Buds Pro - Noise Canceling Earbuds \ldots) \\ \hline \midrule
\textbf{Case \#1: User ID} AHTS2DVIJSO6IARPLRV55RGDHMMA \\ 
\textbf{User profile}: This user consumes electronics primarily to enhance the protection and \textcolor{red}{functionality} of their devices. While they value high quality and durability, they consistently \textcolor{red}{criticize comfort issues and poor functionality}, \ldots.\\ \midrule
\textbf{Item profile:} \\
\textbf{\textit{Audio Technology}}: \textcolor{red}{Active Noise Cancellation} with custom 11 mm drivers \\
\textbf{\textit{Included Accessories}}: Earbuds, Eartips, Wireless Charging Case \\
\textbf{\textit{User-specific Trait}}: Comparative Evaluation; Users are actively comparing these earbuds with competitors, \textcolor{red}{focusing on sound performance, noise cancellation, and overall features.} \ldots \\
\multicolumn{1}{r}{\textbf{$\blacktriangleright$ Rank: 2}} \\ \hline \midrule
\textbf{Case \#2: User ID} AGS4TRMTPU3Y2CMJ52E6DOLX4GPA \\ 
\textbf{User profile}: This consumer primarily seeks electronics that offer 
\textcolor{blue}{seamless compatibility with their devices}, often using them in specific situations such as charging earphones or \textcolor{blue}{operating peripherals with Chromebooks}. \ldots \\ \midrule
\textbf{Our item profile:} \\
\textbf{\textit{Device\slash OS Compatibility}}: \textcolor{blue}{Google Assistant-enabled Android 6.0} \\
\textbf{\textit{Smart and AI Features}}: Hands-free \textcolor{blue}{Google Assistant support}  \\
\textbf{\textit{Store}}: \textcolor{blue}{Google}  \\
\textbf{\textit{User-specific Trait}}: \textcolor{blue}{Brand Loyalty}; There are consumers who exhibit a \textcolor{blue}{strong preference for Google products}, appreciating the familiarity and perceived reliability associated with the brand.  \ldots \\
\multicolumn{1}{r}{\textbf{$\blacktriangleright$ Rank: 1}} \\ \hline \midrule
\textbf{\textit{cf}. EXP3RT item profile:} \\

\textbf{\textit{[Like]}} \\
- \textcolor{red}{Sound quality is improved} compared to the Pixel Buds A Series. \\
- \textcolor{red}{It includes additional features such as ANC and Transparency modes.} \\
- \textcolor{blue}{It works well with other Google products.} \\
- Best stock sound quality of all the earbuds, with excellent touch controls and responsive features (Google Pixelbuds Pro). \ldots \\
\textbf{\textit{[Dislike]}} \\
- It has insecure fit as they do not go into the ear canal or have wingtips for stability.\\
- The case feels cheap and bulky. \\
- It is quite overpriced compared to others (Samsung Galaxy Buds 2 Pro). \ldots \\
\textbf{\textit{[Average Rating]}} 4.7 \\
\multicolumn{1}{r}{\textbf{$\blacktriangleright$ Rank: 19} on Case \#1, \textbf{7} on Case \#2} \\
\bottomrule
\end{tabularx}
\vspace{0.4cm}
\end{table}

\subsubsection{\textbf{Case Study}}

Table~\ref{tab:case} presents item profiles generated by \proposed and EXP3RT for the same item across two different users.
For Case~\#1, who criticizes poor functionality, \proposed selects features around its performance (e.g., noise cancelling).
On the other hand, for Case~\#2, who seeks brand ecosystem compatibility, it instead presents brand-specific information, where the combination of these features collectively reinforces the ecosystem angle to form a coherent insight.
EXP3RT, by contrast, applies a single static profile to both users;
although the resulting profile contains signals relevant to each user, these are buried along with context-irrelevant information, making the LLM struggle to identify what matters for each of the user.
This evidences that user-adaptive profile construction is essential; the same item information can either guide or mislead the LLM depending on how selectively it is presented.

\section{Conclusion}
\label{sec:conclusion}

In this paper, we propose \proposed to address underexplored challenges in item profiling from vast item metadata and item/user context-relevant information.
\proposed first structures raw metadata and reviews into objective features and subjective traits, and employs lightweight selectors to construct user-specialized item profiles with reduced serving-time overhead.
Extensive experiments show that \proposed consistently outperforms existing baselines, supporting that careful organization and user-adaptive selection of item information are as important as providing valuable information.
We expect this study to lay the groundwork for item-side profiling in recommendation, extending beyond the prevailing user-side focus.

\vspace{-0.25cm}

\section*{Acknowledgments}
This work was the result of project supported by KT (Korea Telecom)-Korea University AICT R\&D Center.
This work was also supported by the IITP-ICT Creative Consilience Program grant funded by the MSIT (IITP-2026-RS-2020-II201819), 
Basic Science Research Program through the NRF funded by the Ministry of Education (NRF-2021R1A6A1A03045425), 
the NRF grant funded by the MSIT (RS-2026-25486220), 
and the IITP grant funded by the MSIT (IITP-2026-RS-2026-25616664, AI Star Fellowship Support Program).

\pagebreak

\section*{GenAI Usage Disclosure} 

GPT-4o-mini, Llama3.1, Llama3.2 and Qwen2.5 were used as a methodological component of the proposed framework, as described in the Experiment section.
These models were used only within the reported experimental settings specified in the work.
Generative AI tools (ChatGPT and Claude) were used during manuscript preparation solely for minor language editing, specifically for grammar checking, spelling and typo correction.
These tools were used to suggest debugging directions in code development.
All AI-assisted edits were manually reviewed and verified by the authors before being incorporated into the work.
Apart from the uses disclosed above, no other generative AI tools were used for experimentation, data analysis, or content generation.


\bibliographystyle{ACM-Reference-Format}
\balance
\bibliography{acmart}

@article{recsys-llmera,
  author       = {Zihuai Zhao and
                  Wenqi Fan and
                  Jiatong Li and
                  Yunqing Liu and
                  Xiaowei Mei and
                  Yiqi Wang and
                  Zhen Wen and
                  Fei Wang and
                  Xiangyu Zhao and
                  Jiliang Tang and
                  Qing Li},
  title        = {Recommender Systems in the Era of Large Language Models (LLMs)},
  journal      = {{IEEE} Trans. Knowl. Data Eng.},
  volume       = {36},
  number       = {11},
  pages        = {6889--6907},
  year         = {2024}
}

@inproceedings{mrs-reranking,
  author       = {Weiwen Liu and
                  Jiarui Qin and
                  Ruiming Tang and
                  Bo Chen},
  title        = {Neural Re-ranking for Multi-stage Recommender Systems},
  booktitle    = {RecSys '22: Sixteenth {ACM} Conference on Recommender Systems, Seattle,
                  WA, USA, September 18 - 23, 2022},
  pages        = {698--699},
  publisher    = {{ACM}},
  year         = {2022}
}

@article{llamaRec,
  author       = {Zhenrui Yue and
                  Sara Rabhi and
                  Gabriel de Souza Pereira Moreira and
                  Dong Wang and
                  Even Oldridge},
  title        = {LlamaRec: Two-Stage Recommendation using Large Language Models for
                  Ranking},
  journal      = {CoRR},
  volume       = {abs/2311.02089},
  year         = {2023},
  eprinttype   = {arXiv},
  eprint       = {2311.02089}
}

@article{llm-reranking-1,
  author       = {Diego Carraro and
                  Derek G. Bridge},
  title        = {Enhancing Recommendation Diversity by Re-ranking with Large Language
                  Models},
  journal      = {Trans. Recomm. Syst.},
  volume       = {4},
  number       = {2},
  pages        = {18:1--18:40},
  year         = {2026}
}

@inproceedings{llmrank,
  author       = {Yupeng Hou and
                  Junjie Zhang and
                  Zihan Lin and
                  Hongyu Lu and
                  Ruobing Xie and
                  Julian J. McAuley and
                  Wayne Xin Zhao},
  title        = {Large Language Models are Zero-Shot Rankers for Recommender Systems},
  booktitle    = {Advances in Information Retrieval - 46th European Conference on Information
                  Retrieval, {ECIR} 2024, Glasgow, UK, March 24-28, 2024, Proceedings,
                  Part {II}},
  series       = {Lecture Notes in Computer Science},
  pages        = {364--381},
  publisher    = {Springer},
  year         = {2024}
}

@inproceedings{rec-gpt, 
  author       = {Sunhao Dai and
                  Ninglu Shao and
                  Haiyuan Zhao and
                  Weijie Yu and
                  Zihua Si and
                  Chen Xu and
                  Zhongxiang Sun and
                  Xiao Zhang and
                  Jun Xu},
  title        = {Uncovering ChatGPT's Capabilities in Recommender Systems},
  booktitle    = {Proceedings of the 17th {ACM} Conference on Recommender Systems, RecSys
                  2023, Singapore, Singapore, September 18-22, 2023},
  pages        = {1126--1132},
  publisher    = {{ACM}},
  year         = {2023}
}

@inproceedings{llm4rerank,
  author       = {Jingtong Gao and
                  Bo Chen and
                  Xiangyu Zhao and
                  Weiwen Liu and
                  Xiangyang Li and
                  Yichao Wang and
                  Wanyu Wang and
                  Huifeng Guo and
                  Ruiming Tang},
  title        = {LLM4Rerank: LLM-based Auto-Reranking Framework for Recommendations},
  booktitle    = {Proceedings of the {ACM} on Web Conference 2025, {WWW} 2025, Sydney,
                  NSW, Australia, 28 April 2025- 2 May 2025},
  pages        = {228--239},
  publisher    = {{ACM}},
  year         = {2025}
}

@inproceedings{lm-as-recsys,
  author    = {Zhang, Yuhui and Ding, Hao and Shui, Zeren and Ma, Yifei and Zou, James and Deoras, Anoop and Wang, Hao},
  title     = {Language Models as Recommender Systems: Evaluations and Limitations},
  booktitle = {I (Still) Can't Believe It's Not Better Workshop at NeurIPS},
  year      = {2021}
}

@article{chatgptgoodrecommender,
  author       = {Junling Liu and
                  Chao Liu and
                  Renjie Lv and
                  Kang Zhou and
                  Yan Zhang},
  title        = {Is ChatGPT a Good Recommender? {A} Preliminary Study},
  journal      = {CoRR},
  volume       = {abs/2304.10149},
  year         = {2023},
  eprinttype   = {arXiv},
  eprint       = {2304.10149}
}

@article{palr,
  author       = {Zheng Chen},
  title        = {{PALR:} Personalization Aware LLMs for Recommendation},
  journal      = {CoRR},
  volume       = {abs/2305.07622},
  year         = {2023},
  eprinttype   = {arXiv},
  eprint       = {2305.07622}
}

@inproceedings{llm-trsr,
  author       = {Zhi Zheng and
                  Wenshuo Chao and
                  Zhaopeng Qiu and
                  Hengshu Zhu and
                  Hui Xiong},
  editor       = {Tat{-}Seng Chua and
                  Chong{-}Wah Ngo and
                  Ravi Kumar and
                  Hady W. Lauw and
                  Roy Ka{-}Wei Lee},
  title        = {Harnessing Large Language Models for Text-Rich Sequential Recommendation},
  booktitle    = {Proceedings of the {ACM} on Web Conference 2024, {WWW} 2024, Singapore,
                  May 13-17, 2024},
  pages        = {3207--3216},
  publisher    = {{ACM}},
  year         = {2024}
}

@inproceedings{exp3rt,
  author       = {Jieyong Kim and
                  Hyunseo Kim and
                  Hyunjin Cho and
                  SeongKu Kang and
                  Buru Chang and
                  Jinyoung Yeo and
                  Dongha Lee},
  title        = {Review-driven Personalized Preference Reasoning with Large Language
                  Models for Recommendation},
  booktitle    = {Proceedings of the 48th International {ACM} {SIGIR} Conference on
                  Research and Development in Information Retrieval, {SIGIR} 2025, Padua,
                  Italy, July 13-18, 2025},
  pages        = {1697--1706},
  publisher    = {{ACM}},
  year         = {2025}
}

@inproceedings{langptune,
  author       = {Zhaolin Gao and
                  Joyce Zhou and
                  Yijia Dai and
                  Thorsten Joachims},
  title        = {LangPTune: Optimizing Language-based User Profiles for Recommendation},
  booktitle    = {Proceedings of the 34th {ACM} International Conference on Information
                  and Knowledge Management, {CIKM} 2025, Seoul, Republic of Korea, November
                  10-14, 2025},
  pages        = {707--717},
  publisher    = {{ACM}},
  year         = {2025}
}

@inproceedings{lettingo,
  author       = {Lu Wang and
                  Di Zhang and
                  Fangkai Yang and
                  Pu Zhao and
                  Jianfeng Liu and
                  Yuefeng Zhan and
                  Hao Sun and
                  Qingwei Lin and
                  Weiwei Deng and
                  Dongmei Zhang and
                  Feng Sun and
                  Qi Zhang},
  editor       = {Luiza Antonie and
                  Jian Pei and
                  Xiaohui Yu and
                  Flavio Chierichetti and
                  Hady W. Lauw and
                  Yizhou Sun and
                  Srinivasan Parthasarathy},
  title        = {LettinGo: Explore User Profile Generation for Recommendation System},
  booktitle    = {Proceedings of the 31st {ACM} {SIGKDD} Conference on Knowledge Discovery
                  and Data Mining, V.2, {KDD} 2025, Toronto ON, Canada, August 3-7,
                  2025},
  pages        = {2985--2995},
  publisher    = {{ACM}},
  year         = {2025}
}

@article{languagebaseduserprofilesrecommendation,
  author       = {Joyce Zhou and
                  Yijia Dai and
                  Thorsten Joachims},
  title        = {Language-Based User Profiles for Recommendation},
  journal      = {CoRR},
  volume       = {abs/2402.15623},
  year         = {2024},
  eprinttype   = {arXiv},
  eprint       = {2402.15623}
}

@inproceedings{kar,
  author       = {Yunjia Xi and
                  Weiwen Liu and
                  Jianghao Lin and
                  Xiaoling Cai and
                  Hong Zhu and
                  Jieming Zhu and
                  Bo Chen and
                  Ruiming Tang and
                  Weinan Zhang and
                  Yong Yu},
  title        = {Towards Open-World Recommendation with Knowledge Augmentation from
                  Large Language Models},
  booktitle    = {Proceedings of the 18th {ACM} Conference on Recommender Systems, RecSys
                  2024, Bari, Italy, October 14-18, 2024},
  pages        = {12--22},
  publisher    = {{ACM}},
  year         = {2024}
}

@inproceedings{thinkrec,
  author       = {Qihang Yu and
                  Kairui Fu and
                  Zheqi Lv and
                  Shengyu Zhang and
                  Xinhui Wu and
                  Chen Lin and
                  Feng Wei and
                  Bo Zheng and
                  Fei Wu},
  title        = {ThinkRec: Thinking-based recommendation via {LLM}},
  booktitle    = {Proceedings of the {ACM} Web Conference 2026, {WWW} 2026, Dubai, United
                  Arab Emirates, originally scheduled for April 13-17, 2026, rescheduled
                  for June 29 - July 3, 2026},
  pages        = {5698--5709},
  publisher    = {{ACM}},
  year         = {2026}
}

@article{collm,
  author       = {Yang Zhang and
                  Fuli Feng and
                  Jizhi Zhang and
                  Keqin Bao and
                  Qifan Wang and
                  Xiangnan He},
  title        = {CoLLM: Integrating Collaborative Embeddings Into Large Language Models
                  for Recommendation},
  journal      = {{IEEE} Trans. Knowl. Data Eng.},
  volume       = {37},
  number       = {5},
  pages        = {2329--2340},
  year         = {2025}
}

@inproceedings{adafs,
  author       = {Weilin Lin and
                  Xiangyu Zhao and
                  Yejing Wang and
                  Tong Xu and
                  Xian Wu},
  title        = {AdaFS: Adaptive Feature Selection in Deep Recommender System},
  booktitle    = {{KDD} '22: The 28th {ACM} {SIGKDD} Conference on Knowledge Discovery
                  and Data Mining, Washington, DC, USA, August 14 - 18, 2022},
  pages        = {3309--3317},
  publisher    = {{ACM}},
  year         = {2022}
}

@inproceedings{autofield,
  author       = {Yejing Wang and
                  Xiangyu Zhao and
                  Tong Xu and
                  Xian Wu},
  title        = {AutoField: Automating Feature Selection in Deep Recommender Systems},
  booktitle    = {{WWW} '22: The {ACM} Web Conference 2022, Virtual Event, Lyon, France,
                  April 25 - 29, 2022},
  pages        = {1977--1986},
  publisher    = {{ACM}},
  year         = {2022}
}

@article{amazon-review-dataset-23,
  author       = {Yupeng Hou and
                  Jiacheng Li and
                  Zhankui He and
                  An Yan and
                  Xiusi Chen and
                  Julian J. McAuley},
  title        = {Bridging Language and Items for Retrieval and Recommendation},
  journal      = {CoRR},
  volume       = {abs/2403.03952},
  year         = {2024},
  eprinttype   = {arXiv},
  eprint       = {2403.03952}
}

@inproceedings{m-llm3rec,
  author       = {Lining Chen and
                  Qingwen Zeng and
                  Huaming Chen},
  title        = {M-\emph{LLM\({}^{\mbox{3}}\)}REC: {A} Motivation-Aware User-Item Interaction
                  Framework for Enhancing Recommendation Accuracy with LLMs},
  booktitle    = {Proceedings of the 34th {ACM} International Conference on Information
                  and Knowledge Management, {CIKM} 2025, Seoul, Republic of Korea, November
                  10-14, 2025},
  pages        = {291--300},
  publisher    = {{ACM}},
  year         = {2025}
}

@article{reason4rec,
  author       = {Yi Fang and
                  Wenjie Wang and
                  Yang Zhang and
                  Fengbin Zhu and
                  Qifan Wang and
                  Fuli Feng and
                  Xiangnan He},
  title        = {Reason4Rec: Large Language Models for Recommendation with Deliberative
                  User Preference Alignment},
  journal      = {CoRR},
  volume       = {abs/2502.02061},
  year         = {2025},
  eprinttype   = {arXiv},
  eprint       = {2502.02061}
}

@inproceedings{llmrec,
  author       = {Wei Wei and
                  Xubin Ren and
                  Jiabin Tang and
                  Qinyong Wang and
                  Lixin Su and
                  Suqi Cheng and
                  Junfeng Wang and
                  Dawei Yin and
                  Chao Huang},
  title        = {LLMRec: Large Language Models with Graph Augmentation for Recommendation},
  booktitle    = {Proceedings of the 17th {ACM} International Conference on Web Search
                  and Data Mining, {WSDM} 2024, Merida, Mexico, March 4-8, 2024},
  pages        = {806--815},
  publisher    = {{ACM}},
  year         = {2024}
}

@inproceedings{wide-deep,
  author       = {Heng{-}Tze Cheng and
                  Levent Koc and
                  Jeremiah Harmsen and
                  Tal Shaked and
                  Tushar Chandra and
                  Hrishi Aradhye and
                  Glen Anderson and
                  Greg Corrado and
                  Wei Chai and
                  Mustafa Ispir and
                  Rohan Anil and
                  Zakaria Haque and
                  Lichan Hong and
                  Vihan Jain and
                  Xiaobing Liu and
                  Hemal Shah},
  title        = {Wide {\&} Deep Learning for Recommender Systems},
  booktitle    = {Proceedings of the 1st Workshop on Deep Learning for Recommender Systems,
                  DLRS@RecSys 2016, Boston, MA, USA, September 15, 2016},
  pages        = {7--10},
  publisher    = {{ACM}},
  year         = {2016}
}

@inproceedings{DGDPO,   
  author       = {Hongyang Liu and
                  Zhu Sun and
                  Tianjun Wei and
                  Yan Wang and
                  Jiajie Zhu and
                  Xinghua Qu},
  title        = {Diagnostic-Guided Dynamic Profile Optimization for LLM-based User
                  Simulators in Sequential Recommendation},
  booktitle    = {Fortieth {AAAI} Conference on Artificial Intelligence, Thirty-Eighth
                  Conference on Innovative Applications of Artificial Intelligence,
                  Sixteenth Symposium on Educational Advances in Artificial Intelligence,
                  {AAAI} 2026, Singapore, January 20-27, 2026},
  pages        = {15306--15314},
  publisher    = {{AAAI} Press},
  year         = {2026}}

@inproceedings{deeper,
  author       = {Aili Chen and
                  Chengyu Du and
                  Jiangjie Chen and
                  Jinghan Xu and
                  Yikai Zhang and
                  Siyu Yuan and
                  Zulong Chen and
                  Liangyue Li and
                  Yanghua Xiao},
  title        = {{DEEPER} Insight into Your User: Directed Persona Refinement for Dynamic
                  Persona Modeling},
  booktitle    = {Proceedings of the 63rd Annual Meeting of the Association for Computational
                  Linguistics (Volume 1: Long Papers), {ACL} 2025, Vienna, Austria,
                  July 27 - August 1, 2025},
  pages        = {24157--24180},
  publisher    = {Association for Computational Linguistics},
  year         = {2025}
}

@article{lostinthemiddle,
  author       = {Nelson F. Liu and
                  Kevin Lin and
                  John Hewitt and
                  Ashwin Paranjape and
                  Michele Bevilacqua and
                  Fabio Petroni and
                  Percy Liang},
  title        = {Lost in the Middle: How Language Models Use Long Contexts},
  journal      = {Trans. Assoc. Comput. Linguistics},
  volume       = {12},
  pages        = {157--173},
  year         = {2024}
}

@inproceedings{tallrec,
  author       = {Keqin Bao and
                  Jizhi Zhang and
                  Yang Zhang and
                  Wenjie Wang and
                  Fuli Feng and
                  Xiangnan He},
  title        = {TALLRec: An Effective and Efficient Tuning Framework to Align Large
                  Language Model with Recommendation},
  booktitle    = {Proceedings of the 17th {ACM} Conference on Recommender Systems, RecSys
                  2023, Singapore, Singapore, September 18-22, 2023},
  pages        = {1007--1014},
  publisher    = {{ACM}},
  year         = {2023}
}

@inproceedings{bpr,
  author       = {Steffen Rendle and
                  Christoph Freudenthaler and
                  Zeno Gantner and
                  Lars Schmidt{-}Thieme},
  title        = {{BPR:} Bayesian Personalized Ranking from Implicit Feedback},
  booktitle    = {{UAI} 2009, Proceedings of the Twenty-Fifth Conference on Uncertainty
                  in Artificial Intelligence, Montreal, QC, Canada, June 18-21, 2009},
  pages        = {452--461},
  publisher    = {{AUAI} Press},
  year         = {2009}
}

@inproceedings{sasrec,
  author       = {Wang{-}Cheng Kang and
                  Julian J. McAuley},
  title        = {Self-Attentive Sequential Recommendation},
  booktitle    = {{IEEE} International Conference on Data Mining, {ICDM} 2018, Singapore,
                  November 17-20, 2018},
  pages        = {197--206},
  publisher    = {{IEEE} Computer Society},
  year         = {2018}
}

@inproceedings{bert4rec,
  author       = {Fei Sun and
                  Jun Liu and
                  Jian Wu and
                  Changhua Pei and
                  Xiao Lin and
                  Wenwu Ou and
                  Peng Jiang},
  title        = {BERT4Rec: Sequential Recommendation with Bidirectional Encoder Representations
                  from Transformer},
  booktitle    = {Proceedings of the 28th {ACM} International Conference on Information
                  and Knowledge Management, {CIKM} 2019, Beijing, China, November 3-7,
                  2019},
  pages        = {1441--1450},
  publisher    = {{ACM}},
  year         = {2019}
}

@article{bge-m3,
  author       = {Jianlv Chen and
                  Shitao Xiao and
                  Peitian Zhang and
                  Kun Luo and
                  Defu Lian and
                  Zheng Liu},
  title        = {{BGE} M3-Embedding: Multi-Lingual, Multi-Functionality, Multi-Granularity
                  Text Embeddings Through Self-Knowledge Distillation},
  journal      = {CoRR},
  volume       = {abs/2402.03216},
  year         = {2024},
  eprinttype   = {arXiv},
  eprint       = {2402.03216}
}

@inproceedings{recsaver,
  author       = {Alicia Tsai and
                  Adam Kraft and
                  Long Jin and
                  Chenwei Cai and
                  Anahita Hosseini and
                  Taibai Xu and
                  Zemin Zhang and
                  Lichan Hong and
                  Ed Huai{-}hsin Chi and
                  Xinyang Yi},
  title        = {Leveraging {LLM} Reasoning Enhances Personalized Recommender Systems},
  booktitle    = {Findings of the Association for Computational Linguistics, {ACL} 2024,
                  Bangkok, Thailand and virtual meeting, August 11-16, 2024},
  series       = {Findings of {ACL}},
  pages        = {13176--13188},
  publisher    = {Association for Computational Linguistics},
  year         = {2024}
}

@inproceedings{weaverec,
  author       = {Min Hou and
                  Xin Liu and
                  Le Wu and
                  Chenyi He and
                  Hao Liu and
                  Zhi Li and
                  Xin Li and
                  Si Wei},
  title        = {WeaveRec: An LLM-Based Cross-Domain Sequential Recommendation Framework
                  with Model Merging},
  booktitle    = {Proceedings of the {ACM} Web Conference 2026, {WWW} 2026, Dubai, United
                  Arab Emirates, originally scheduled for April 13-17, 2026, rescheduled
                  for June 29 - July 3, 2026},
  pages        = {6342--6353},
  publisher    = {{ACM}},
  year         = {2026}
}

@inproceedings{msl,
  author       = {Bohao Wang and
                  Feng Liu and
                  Jiawei Chen and
                  Xingyu Lou and
                  Changwang Zhang and
                  Jun Wang and
                  Yuegang Sun and
                  Yan Feng and
                  Chun Chen and
                  Can Wang},
  title        = {{MSL:} Not All Tokens Are What You Need for Tuning {LLM} as a Recommender},
  booktitle    = {Proceedings of the 48th International {ACM} {SIGIR} Conference on
                  Research and Development in Information Retrieval, {SIGIR} 2025, Padua,
                  Italy, July 13-18, 2025},
  pages        = {1912--1922},
  publisher    = {{ACM}},
  year         = {2025}
}

@inproceedings{netflix-llm-reasoning,
  author       = {Shijun Li and
                  Yu Wang and
                  Jin Wang and
                  Ying Li and
                  Joydeep Ghosh and
                  Anne Cocos},
  title        = {{LLM} Reasoning for Cold-Start Item Recommendation},
  booktitle    = {Proceedings of the {ACM} Web Conference 2026, {WWW} 2026, Dubai, United
                  Arab Emirates, originally scheduled for April 13-17, 2026, rescheduled
                  for June 29 - July 3, 2026},
  pages        = {8409--8412},
  publisher    = {{ACM}},
  year         = {2026}
}

@inproceedings{P5, 
  author       = {Shijie Geng and
                  Shuchang Liu and
                  Zuohui Fu and
                  Yingqiang Ge and
                  Yongfeng Zhang},
  title        = {Recommendation as Language Processing {(RLP):} {A} Unified Pretrain,
                  Personalized Prompt {\&} Predict Paradigm {(P5)}},
  booktitle    = {RecSys '22: Sixteenth {ACM} Conference on Recommender Systems, Seattle,
                  WA, USA, September 18 - 23, 2022},
  pages        = {299--315},
  publisher    = {{ACM}},
  year         = {2022}
}

@inproceedings{upr,
  author       = {Jerome Ramos and
                  Hossein A. Rahmani and
                  Xi Wang and
                  Xiao Fu and
                  Aldo Lipani},
  title        = {Transparent and Scrutable Recommendations Using Natural Language User
                  Profiles},
  booktitle    = {Proceedings of the 62nd Annual Meeting of the Association for Computational
                  Linguistics (Volume 1: Long Papers), {ACL} 2024, Bangkok, Thailand,
                  August 11-16, 2024},
  pages        = {13971--13984},
  publisher    = {Association for Computational Linguistics},
  year         = {2024}
}

@inproceedings{xdeepfm,
  author       = {Jianxun Lian and
                  Xiaohuan Zhou and
                  Fuzheng Zhang and
                  Zhongxia Chen and
                  Xing Xie and
                  Guangzhong Sun},
  title        = {xDeepFM: Combining Explicit and Implicit Feature Interactions for
                  Recommender Systems},
  booktitle    = {Proceedings of the 24th {ACM} {SIGKDD} International Conference on
                  Knowledge Discovery {\&} Data Mining, {KDD} 2018, London, UK,
                  August 19-23, 2018},
  pages        = {1754--1763},
  publisher    = {{ACM}},
  year         = {2018}
}

@article{qwen2,
  author       = {An Yang and
                  Baosong Yang and
                  Beichen Zhang and
                  Binyuan Hui and
                  Bo Zheng and
                  Bowen Yu and
                  Chengyuan Li and
                  Dayiheng Liu and
                  Fei Huang and
                  Haoran Wei and
                  Huan Lin and
                  Jian Yang and
                  Jianhong Tu and
                  Jianwei Zhang and
                  Jianxin Yang and
                  Jiaxi Yang and
                  Jingren Zhou and
                  Junyang Lin and
                  Kai Dang and
                  Keming Lu and
                  Keqin Bao and
                  Kexin Yang and
                  Le Yu and
                  Mei Li and
                  Mingfeng Xue and
                  Pei Zhang and
                  Qin Zhu and
                  Rui Men and
                  Runji Lin and
                  Tianhao Li and
                  Tingyu Xia and
                  Xingzhang Ren and
                  Xuancheng Ren and
                  Yang Fan and
                  Yang Su and
                  Yichang Zhang and
                  Yu Wan and
                  Yuqiong Liu and
                  Zeyu Cui and
                  Zhenru Zhang and
                  Zihan Qiu},
  title        = {Qwen2.5 Technical Report},
  journal      = {CoRR},
  volume       = {abs/2412.15115},
  year         = {2024},
  eprinttype   = {arXiv},
  eprint       = {2412.15115}
}

@misc{llama3,
  author       = {Llama Team},
  title        = {The Llama 3 Herd of Models},
  journal      = {CoRR},
  volume       = {abs/2407.21783},
  year         = {2024},
  eprinttype   = {arXiv},
  eprint       = {2407.21783}
}

@article{ali-rerank,
  author       = {Guangda Huzhang and
                  Zhen{-}Jia Pang and
                  Yongqing Gao and
                  Yawen Liu and
                  Weijie Shen and
                  Wen{-}Ji Zhou and
                  Qianying Lin and
                  Qing Da and
                  Anxiang Zeng and
                  Han Yu and
                  Yang Yu and
                  Zhi{-}Hua Zhou},
  title        = {AliExpress Learning-to-Rank: Maximizing Online Model Performance Without
                  Going Online},
  journal      = {{IEEE} Trans. Knowl. Data Eng.},
  volume       = {35},
  number       = {2},
  pages        = {1214--1226},
  year         = {2023}
  }

@inproceedings{kuairand-dataset,
  author       = {Chongming Gao and
                  Shijun Li and
                  Yuan Zhang and
                  Jiawei Chen and
                  Biao Li and
                  Wenqiang Lei and
                  Peng Jiang and
                  Xiangnan He},
  title        = {KuaiRand: An Unbiased Sequential Recommendation Dataset with Randomly
                  Exposed Videos},
  booktitle    = {Proceedings of the 31st {ACM} International Conference on Information
                  {\&} Knowledge Management, Atlanta, GA, USA, October 17-21, 2022},
  pages        = {3953--3957},
  publisher    = {{ACM}},
  year         = {2022}
}

@inproceedings{yahoo-news-dataset,
  author       = {Lihong Li and
                  Wei Chu and
                  John Langford and
                  Robert E. Schapire},
  title        = {A contextual-bandit approach to personalized news article recommendation},
  booktitle    = {Proceedings of the 19th International Conference on World Wide Web,
                  {WWW} 2010, Raleigh, North Carolina, USA, April 26-30, 2010},
  pages        = {661--670},
  publisher    = {{ACM}},
  year         = {2010}
}

@inproceedings{reaction, 
  author       = {Song{-}Li Wu and
                  Zhaocheng Du and
                  Qinglin Jia and
                  Zhenhua Dong},
  title        = {{REACTION:} Parameter-Efficient Learning for Recommendation},
  booktitle    = {Fortieth {AAAI} Conference on Artificial Intelligence, Thirty-Eighth
                  Conference on Innovative Applications of Artificial Intelligence,
                  Sixteenth Symposium on Educational Advances in Artificial Intelligence,
                  {AAAI} 2026, Singapore, January 20-27, 2026},
  pages        = {15977--15985},
  publisher    = {{AAAI} Press},
  year         = {2026}
}

@inproceedings{reap,
  author       = {Yijie Zhu and
                  Haojie Zhou and
                  Wanting Hong and
                  Tailin Liu and
                  Ning Wang},
  title        = {{REAP:} Enhancing {RAG} with Recursive Evaluation and Adaptive Planning
                  for Multi-Hop Question Answering},
  booktitle    = {Fortieth {AAAI} Conference on Artificial Intelligence, Thirty-Eighth
                  Conference on Innovative Applications of Artificial Intelligence,
                  Sixteenth Symposium on Educational Advances in Artificial Intelligence,
                  {AAAI} 2026, Singapore, January 20-27, 2026},
  pages        = {35230--35238},
  publisher    = {{AAAI} Press},
  year         = {2026}
}

@inproceedings{hadsf,
  author       = {Zheng Nie and
                  Peijie Sun},
  title        = {{HADSF:} Aspect Aware Semantic Control for Explainable Recommendation},
  booktitle    = {Proceedings of the Nineteenth {ACM} International Conference on Web
                  Search and Data Mining, {WSDM} 2026, Boise, ID, USA, February 22-26,
                  2026},
  pages        = {509--519},
  publisher    = {{ACM}},
  year         = {2026}
}

@inproceedings{tgd-prompt-opt,
  author       = {Reid Pryzant and
                  Dan Iter and
                  Jerry Li and
                  Yin Tat Lee and
                  Chenguang Zhu and
                  Michael Zeng},
  title        = {Automatic Prompt Optimization with "Gradient Descent" and Beam Search},
  booktitle    = {Proceedings of the 2023 Conference on Empirical Methods in Natural
                  Language Processing, {EMNLP} 2023, Singapore, December 6-10, 2023},
  pages        = {7957--7968},
  publisher    = {Association for Computational Linguistics},
  year         = {2023}
}

@inproceedings{math-shepherd,
  author       = {Peiyi Wang and
                  Lei Li and
                  Zhihong Shao and
                  Runxin Xu and
                  Damai Dai and
                  Yifei Li and
                  Deli Chen and
                  Yu Wu and
                  Zhifang Sui},
  title        = {Math-Shepherd: Verify and Reinforce LLMs Step-by-step without Human
                  Annotations},
  booktitle    = {Proceedings of the 62nd Annual Meeting of the Association for Computational
                  Linguistics (Volume 1: Long Papers), {ACL} 2024, Bangkok, Thailand,
                  August 11-16, 2024},
  pages        = {9426--9439},
  publisher    = {Association for Computational Linguistics},
  year         = {2024}
}

@inproceedings{verify-step,
  author       = {Hunter Lightman and
                  Vineet Kosaraju and
                  Yuri Burda and
                  Harrison Edwards and
                  Bowen Baker and
                  Teddy Lee and
                  Jan Leike and
                  John Schulman and
                  Ilya Sutskever and
                  Karl Cobbe},
  title        = {Let's Verify Step by Step},
  booktitle    = {The Twelfth International Conference on Learning Representations,
                  {ICLR} 2024, Vienna, Austria, May 7-11, 2024},
  publisher    = {OpenReview.net},
  year         = {2024}
}

@inproceedings{kg-llm-rec,
  author       = {Shijie Wang and
                  Wenqi Fan and
                  Yue Feng and
                  Shanru Lin and
                  Xinyu Ma and
                  Shuaiqiang Wang and
                  Dawei Yin},
  title        = {Knowledge Graph Retrieval-Augmented Generation for LLM-based Recommendation},
  booktitle    = {Proceedings of the 63rd Annual Meeting of the Association for Computational
                  Linguistics (Volume 1: Long Papers), {ACL} 2025, Vienna, Austria,
                  July 27 - August 1, 2025},
  pages        = {27152--27168},
  publisher    = {Association for Computational Linguistics},
  year         = {2025}
}

@article{agentic-tagger,
  author       = {Zhouhang Xie and
                  Bo Peng and
                  Zhankui He and
                  Ziqi Chen and
                  Alice Han and
                  Isabella Ye and
                  Benjamin Coleman and
                  Noveen Sachdeva and
                  Fernando Pereira and
                  Julian J. McAuley and
                  Wang{-}Cheng Kang and
                  Derek Zhiyuan Cheng and
                  Beidou Wang and
                  Randolph Brown},
  title        = {AgenticTagger: Structured Item Representation for Recommendation with
                  {LLM} Agents},
  journal      = {CoRR},
  volume       = {abs/2602.05945},
  year         = {2026},
  eprinttype   = {arXiv},
  eprint       = {2602.05945}
}

@inproceedings{uncertainty-quantification-llm-rec,
  author       = {Wonbin Kweon and
                  Sanghwan Jang and
                  SeongKu Kang and
                  Hwanjo Yu},
  title        = {Uncertainty Quantification and Decomposition for LLM-based Recommendation},
  booktitle    = {Proceedings of the {ACM} on Web Conference 2025, {WWW} 2025, Sydney,
                  NSW, Australia, 28 April 2025- 2 May 2025},
  pages        = {4889--4901},
  publisher    = {{ACM}},
  year         = {2025}
}

@inproceedings{filling-the-gaps,
  author       = {Jaehyun Lee and
                  Sanghwan Jang and
                  Seongku Kang and
                  Hwanjo Yu},
  title        = {Filling the Gaps: Selective Knowledge Augmentation for {LLM} Recommenders},
  booktitle    = {Proceedings of the 49th International {ACM} {SIGIR} Conference on
                  Research and Development in Information Retrieval, {SIGIR} 2026, MelbourneVICAustralia,
                  July 20-24, 2026},
  pages        = {891--901},
  publisher    = {{ACM}},
  year         = {2026},
}

@inproceedings{mvfs,
  author       = {Youngjune Lee and
                  Yeongjong Jeong and
                  Keunchan Park and
                  SeongKu Kang},
  title        = {MvFS: Multi-view Feature Selection for Recommender System},
  booktitle    = {Proceedings of the 32nd {ACM} International Conference on Information
                  and Knowledge Management, {CIKM} 2023, Birmingham, United Kingdom,
                  October 21-25, 2023},
  pages        = {4048--4052},
  publisher    = {{ACM}},
  year         = {2023}
}

@inproceedings{sprint,
  author       = {Gyuseok Lee and
                  Wonbin Kweon and
                  Zhenrui Yue and
                  Yaokun Liu and
                  Yifan Liu and
                  Susik Yoon and
                  Dong Wang and
                  Seongku Kang},
  title        = {{SPRINT:} Scalable and Predictive Intent Refinement for LLM-Enhanced
                  Session-based Recommendation},
  booktitle    = {Proceedings of the 49th International {ACM} {SIGIR} Conference on
                  Research and Development in Information Retrieval, {SIGIR} 2026, MelbourneVICAustralia,
                  July 20-24, 2026},
  pages        = {902--912},
  publisher    = {{ACM}},
  year         = {2026}
}

@inproceedings{corank,
  author       = {Runchu Tian and
                  Xueqiang Xu and
                  Bowen Jin and
                  SeongKu Kang and
                  Jiawei Han},
  title        = {LLM-Based Compact Reranking with Document Features for Scientific
                  Retrieval},
  booktitle    = {Proceedings of the 32st {ACM} {SIGKDD} Conference on Knowledge Discovery
                  and Data Mining, {KDD} 2026, Jeju, Korea, August 9-13,
                  2026},
  pages        = {12114--12125},
  publisher    = {{ACM}},
  year         = {2026}
}


\newpage

\end{document}